\documentclass[11pt,thmsa,a4paper]{article}
\usepackage{latexsym}
\usepackage{graphics}
\usepackage[dvips]{graphicx}
\usepackage{tabularx,multirow,rotating}
\usepackage{natbib}
\usepackage{mathrsfs}
\usepackage{fancyhdr}
\usepackage{caption}
\usepackage{subfigure}
\usepackage{marvosym}
\usepackage{lscape}
\usepackage{ragged2e}
\usepackage{mathrsfs}
\usepackage{amsfonts}
\usepackage{colortbl}
\usepackage{slashbox}
\usepackage{amsmath,amssymb,fancybox}
\usepackage{hyperref}
\usepackage{rotating}
\usepackage{float}
\hypersetup{
	anchorcolor=black,
	citecolor=blue,
	linkcolor=black,
	filecolor=magenta, 
	urlcolor=red,
	runcolor=black,
	menucolor=black,
	filecolor=black,
	backref=true,
	colorlinks=true,
	pdfpagelabels,
	plainpages=false,
	pdfauthor={},
	pdftitle={},
	pdfkeywords={},
	pdfsubject={},
	pdfstartview=FitH
}
\usepackage{booktabs}
\usepackage{bbm}
\usepackage{resizegather}
\usepackage{footnote}
\usepackage[latin1]{inputenc}
\usepackage[english]{babel}
\usepackage{times}
\usepackage{color}
\usepackage{epsfig}
\usepackage{mathtools}

\usepackage{enumitem}
\newcommand{\subscript}[2]{$#1 _ #2$}
\usepackage{epstopdf}
\usepackage{pgf}
\usepackage{eucal}
\usepackage[T1]{fontenc}
\usepackage{amsthm}
\usepackage{setspace}
\usepackage{natbib}
\usepackage{float}
\usepackage{amsfonts}
\usepackage{graphicx}
\usepackage{adjustbox}
\usepackage{stackrel}
\usepackage{cleveref}
\newtheorem{theorem}{Theorem}

\newtheorem{definition}{Definition}
\newtheorem{assump}{Assumption}
\newenvironment{myassump}[2][]
{\renewcommand\theassump{#2}\begin{assump}[#1]}
	{\end{assump}}
\providecommand{\U}[1]{\protect\rule{.1in}{.1in}}
\begin{document}

\title{Wavelet Estimation for Factor Models with Time-Varying Loadings}
\author{Duv\'an Humberto Cata\~no\footnote{The author would like to thank financial supports from CAPES, CNPq, and University of Sao Paolo, Brazil.} \thanks{University of Antioquia, Colombia.},\\
 C.	 Vladimir Rodr\'iguez-Caballero\thanks{Corresponding author. Department of Statistics, ITAM, Mexico \& CREATES, Aarhus University, Denmark. Address: R\'io Hondo No.1, Col. Progreso Tizap\'an, \'Alvaro Obreg\'on, CDMX. 01080. Mexico. E-mail: vladimir.rodriguez@itam.mx}, \\ 
 Chang Chiann \thanks{University of Sao Paulo, IME, S\~ao Paulo, Brasil}, \\
Daniel Pe\~na \thanks{Department of Statistics and Institute UC3M-BS of Financial Big Data. Universidad Carlos III de Madrid, Getafe, Spain.}
}
\date{}
\maketitle

\begin{abstract}
We introduce a high-dimensional factor model with time-varying loadings. We cover both stationary and nonstationary factors to increase the possibilities of applications. We propose an estimation procedure based on two stages. First, we estimate common factors by principal components. In the second step, considering the estimated factors as observed, the time-varying loadings are estimated by an iterative generalized least squares procedure using wavelet functions. We investigate the finite sample features by some Monte Carlo simulations. Finally, we apply the model to study the Nord Pool power market's electricity prices and loads. \newline


\end{abstract}

\section{Introduction}

Factor models have been widely used in the last decade due to their
ability to explain the structure of a common variability among time series
by a small number of unobservable common factors. In this sense, these
models are used to reduce the dimensionality of complex systems. The studies of
factor models have encompassed the stationary and nonstationary frameworks
by different estimation methods, see among many others, \cite{pena1987identifying}, 
\cite{forni2000generalized}, \cite{stock2002forecasting}, \cite%
{bai2002determining}, \cite{bai2003inferential}, \cite{forni2004generalized}%
, \cite{forni2005generalized}, \cite{lam2012factor}, \cite{bai2016maximum}, \cite{chan2017factor}, and \cite{gao2019structural} for stationary cases, and \cite{bai2004panic}, \cite{pena2006nonstationary}, \cite{barigozzi2016non}, and \cite{rodriguez2017estimation} for nonstationary cases.

The literature has recently focused on a particular framework of factor models that allow the loadings to vary over time. \cite{motta2011locally} introduce
deterministic smooth variations in factor loadings and propose estimation
procedures based on locally weighted generalized least squares using kernel
functions under a time-domain approach. Similarly, \cite{eichler2011fitting} allow for a nonstationary structure in the factor model with deterministic time-dependent functions in factor loadings. Their estimation procedure can be seen as a time-varying spectral density matrix of the underlying process. Furthermore, \cite{mikkelsen2018consistent}
propose a factor model with time-varying loadings that evolve as stationary
VAR processes employing Kalman filter procedures to obtain the maximum
likelihood estimators of the factor loadings parameters.

In this paper, we use wavelet functions to define smooth variations of
loadings in high-dimensional factor models. Our model can be helpful
in some economic applications when the dynamics are driven by smooth variations, whose cumulative effects cannot be ignored,
see \cite{suwang2017}, and \cite{baihan2016}, for details. In our framework,
to capture the common smooth variations in the vector of time series, 
parameters in the loading matrix are assumed to be well approximated by
deterministic functions over time.

The estimation procedure consists of two steps. First, following  \cite{mikkelsen2018consistent}, common factors
are estimated by principal component analysis (PCA). In the
second step, using estimated factors of the first stage, factor loadings are
estimated by an iterative procedure that combines generalized least squares
(GLS) using wavelet functions. We show that factors estimated by
principal components are consistent after controlling the magnitude of the
loadings' instabilities. We highlight that a requirement for such consistency is that common factors
need to be independent of factor loadings' functions. We also study the case of nonstationary factors following the 
approach of \cite{pena2006nonstationary}, who propose to use a generalized covariance matrix to estimate common factors.

We use Monte Carlo simulations to
show that the method correctly identifies the factors and loadings even in relatively small samples. Finally,
we use the methodology proposed to analyze the Nord Pool power market. We use the electricity
system prices and loads throughout two years and a half to show that factor
loadings are not invariant along time, illustrating our model's usefulness. We find that some features of electricity
prices and loads, see, e.g., \cite{weron2007modeling}, and \cite{Weron20141030}, are well extracted by common factors estimates and by time-varying loadings estimates.

In short, the contributions of this article are i) We propose wavelet functions to estimate time-varying loadings and a consistent method to estimate them; ii) we also allow for nonstationary factors to cover more possibilities of application, and iii) the usefulness of the model is discussed through a Monte Carlo study, and by the empirical application. 

The remainder of the paper is organized as follows. Section 2 introduces the
model, shows the consistency of principal components with stationary and
nonstationary variables, and introduces the wavelet functions. Section 3
discusses the estimation procedure. Monte Carlo experiments
are presented in Section 4, whereas an empirical illustration is provided in
Section 5. Finally, Section 6 concludes. We use bold-unslanted
letters for matrices, bold-slanted letters for vectors, and unbold 
letters for scalars. We denote by $tr(\cdot )$ the trace operator, by $rank(%
\mathbf{A})$ the rank of a matrix $\mathbf{A}$, by $\mathbb{I}_{n}$ the
identity matrix of dimension $n$, by $\otimes $ the Kronecker product and by 
$\Vert \cdot \Vert $ the Frobenius (Euclidean) norm, i.e., $\Vert \mathbf{A}%
\Vert =\sqrt{tr(\mathbf{A}^{\prime }\mathbf{A})}.$

\section{The model}
In this section, we introduce the model. First, we consider stationary factors to motivate the general setup and the estimation procedure. Then, in the next section, we relax the stationarity assumption to allow for nonstationary factors.

 \subsection{Stationary factors}
 
The model we propose is as follows
\begin{eqnarray}
\mathbf{Y}_{t} &=&\mathbf{X}_{t}+\mathbf{e}_{t},  \label{3Mod} \\
\mathbf{X}_{t} &=&[\boldsymbol{\Lambda }_{0}+\boldsymbol{\Lambda }(t/T)]%
\mathbf{F}_{t},  \label{3Mod2}
\end{eqnarray}%
where the common component, $\textbf{X}_t$, is an $N-$dimensional  locally stationary process, in the sense of \cite{dahlhaus1997fitting}, and the loadings are defined in the re-scaled time, $u=t/T\in [0,1]$. 
$\textbf{F}_t$ are the unobservable common factors and $\textbf{e}_t$, the idiosyncratic component, is a sequence of weakly dependent variables, see e.g. \cite{stock2002forecasting}, and \cite{bai2003inferential}.

$\boldsymbol{\Lambda}(u)=\{\lambda_{ij}(u), \ \ i=1,\ldots,N; \ \ j=1,\ldots,r\}$, is the  time-dependent factor loading matrix. In this respect, the influence of $\textbf{F}_t$ on the observed process varies over time, and  
$\boldsymbol{\Lambda}(u)$ captures the smooth variations of the loadings with respect $\boldsymbol{\Lambda}_0$, a constant loading matrix.  

When $\boldsymbol{\Lambda}(u)=0, \ \  \forall{u\in[0,1]}$, the standard factor model is considered; therefore, the model proposed in (\ref{3Mod}) can be seen as a generalization of the standard factor model of \cite{bai2003inferential}.
\begin{definition}
\label{def4} The sequence $\mathbf{Y}_t$ in (\ref{3Mod}) follows a
factor model with time-varying loadings if:
\begin{enumerate}
\item[a.] For $N\in \mathbb{N}$, there is a function with 
\begin{equation*}
\begin{array}{ccccc}
\boldsymbol{\Lambda }(\cdot ) & : & [0,1] & \rightarrow  & \mathbb{R}%
^{N\times r} \\ 
&  & u & \mapsto  & \boldsymbol{\Lambda }(u),%
\end{array}%
\end{equation*}%
such that $\forall\; T \in \mathbb{N},$ 
\begin{equation*}
\Gamma _{Y}(u)=[\boldsymbol{\Lambda }_{0}+\boldsymbol{\Lambda }(u)]\Gamma
_{F}[\boldsymbol{\Lambda }_{0}+\boldsymbol{\Lambda }(u)]^{\prime }+\Gamma
_{e},
\end{equation*}%
with $rank[\boldsymbol{\Lambda }_{0}+\boldsymbol{\Lambda }(u)]=r,\quad $ and 
$\quad \Gamma _{F}=\mathbb{V}ar(\mathbf{F}_{t})$ is a positive definite
diagonal matrix.
\item[b.] $\Gamma _{e}=var(\mathbf{e}_{t})$ is a positive definite matrix.
\end{enumerate}
\end{definition}

With an arbitrary constant $Q\in \mathbb{R}^{+}$, we then use the following assumptions to study the model in (\ref{3Mod})  following \cite{bai2002determining}.

\begin{myassump}{A} \textbf{Factors:} 
	\begin{enumerate}[label=\subscript{A}{\arabic*}]
		\item $\mathbb{E}\| \textbf{F}_t \|^4 < Q,$
		\item $T^{-1}\sum_{t=1}^{T}\mathbf{F}_{t}\mathbf{F}_{t}^{'}\stackbin[]{p}\rightarrow \Gamma_F$ as $T\rightarrow\infty,$ with $\Gamma_F$ is a positive definite diagonal matrix.
	\end{enumerate} 
\end{myassump}

\begin{myassump}{B} \textbf{Factor loadings:}
	\begin{enumerate}[label=\subscript{B}{\arabic*}]
	\item $\Vert \boldsymbol{\lambda}_{i0}\Vert \leq \bar{\lambda}$ and $\Vert \boldsymbol{\Lambda }_{0}^{^{\prime }}\boldsymbol{%
\Lambda }_{0}/N-D\Vert \rightarrow 0$, as $N\rightarrow \infty $, with $%
r\times r$ positive definite matrix $D$, and 
$\boldsymbol{\lambda}_{i0}$ is the $i-$th row of $\boldsymbol{\Lambda }_{0},$

\item $\sup_{u\in (0,1)}{\Vert \boldsymbol{\lambda}_{i}(u)\Vert }\leq \bar{\lambda}%
<\infty,$ where 
$\boldsymbol{\lambda}_{i}(u)$ is the $i-$th row of $\boldsymbol{\Lambda }(u),$
\item $\lambda _{ij}(u)\in L^{2}[0,1]$, \quad for $i=1,\ldots ,N$ and $j=1,\ldots
,r.$
\end{enumerate}
\end{myassump}

\begin{myassump}{C} \textbf{Idiosyncratic terms:} 
	\begin{enumerate}[label=\subscript{C}{\arabic*}]
\item $\mathbb{E}(e_{it})=0, \mathbb{E}|e_{it}|^{4}<Q,$
\item $\mathbb{E}[e_{it}e_{jt}]=\tau _{ij,t},$ with $|\tau _{ij,t}|<|\tau_{ij}|$,  for a constant $|\tau _{ij}|$ and $N^{-1}\sum_{i,j=1}^{N}|\tau _{ij}|<Q, \forall {t},$ 
\item $\mathbb{E}[N^{-1}\sum_{i=1}^Ne_{is}e_{it}]=\gamma_N(s,t),$ 
$|\gamma_N(s,s)|<Q, \forall (s,t);$ and \\
$T^{-1}\sum_{s=1}^T\sum_{t=1}^T|\gamma_N(s,t)|<Q.$
\end{enumerate}
\end{myassump}

\begin{myassump}{D} \textbf{Time-varying factor loadings and factors:}\\ 
	$K_{1NT}$, $K_{2NT}$ and $K_{3NT}$ are functions such that the following conditions are fulfilled for any $n,m,k,q=1,\ldots,r$ and $\lambda_{ij}(s/T)\equiv\lambda_{ij}(s).$ 
	\begin{enumerate}[label=\subscript{D}{\arabic*}]
\item $\sup_{s,t}\sum_{i,j}^{N}|\lambda_{in}(s)\lambda_{jm}(t)||\mathbb{E}%
[F_{ns}F_{mt}]|<K_{1NT}$,
\item $\sum_{s,t=1}^{T}\sum_{i,j=1}^{N}|\lambda_{in}(s)\lambda_{jm}(s)||%
\mathbb{E}[F_{ns}F_{ms}F_{kt}F_{qt}]|<K_{2NT},$
\item $\sup_{s,t}\sum_{s=1}^{T}\sum_{i,j=1}^{N}|\lambda_{in}(s)\lambda_{jm}(s)%
\lambda_{ik}(t)\lambda_{jq}(t)||\mathbb{E}%
[F_{ns}F_{ms}F_{kt}F_{qt}]|<K_{3NT} $.
\end{enumerate}
\end{myassump}

\begin{myassump}{E} \textbf{Independence:}\\ 
	The process $e_{it}$ and $F_{js}$ are independent of each other for any $(i,j,s,t)$.
\end{myassump}

Assumption A imposes standard moment conditions. The unobservable
factors have finite fourth moments, and their covariance converges in
probability to a positive definite matrix. Assumptions $B_1$ and $B_2$
ensure that each factor has a nontrivial contribution to the variance of $%
\mathbf{Y}_t$. Assumption $B_3$ ensures the existence of the expansion in the
wavelet function for the factor loadings. Assumption C allows 
dependence in the idiosyncratic process. Assumption D is
required to guarantee the consistency of the principal components. Finally,
independence between factors and the idiosyncratic term is provided in
Assumption E.

\subsubsection{Principal component estimators}

\label{FHAT} We use PCA to estimate the factors $\mathbf{F}_t$. It is
well-known that principal components are obtained solving the optimization
problem 
\begin{equation}  \label{ECPP}
\min_{\mathbf{F},\boldsymbol{\Lambda}}(NT)^{-1}\sum_{i=1}^{N}%
\sum_{t=1}^{T}(Y_{it}-\boldsymbol{\lambda}_i^{\prime }\mathbf{F}_t)^2,
\end{equation}
where $\mathbf{F} = (\mathbf{F}_1, \mathbf{F}_2, \ldots, \mathbf{F}%
_T)^{\prime }$ is a $T\times r$ matrix and ${\boldsymbol{\Lambda}}$ is a $%
N\times r$ matrix. We need to impose some restrictions to guarantee the
identification of the parameters. Solving for $\boldsymbol{\Lambda}$%
, the normalization $\mathbf{F}^{\prime }\mathbf{F}=\mathbb{I}_r$ provides
the necessary number of restrictions. 

With this, minimize (\ref{ECPP}) is equivalent to maximize $tr[\mathbf{F}%
^{\prime }(\mathbf{Y}\mathbf{Y}^{\prime })\mathbf{F}]$, where $\mathbf{Y}=(%
\mathbf{Y}_1, \ldots, \mathbf{Y}_T)^{\prime }$. Then, the estimated factor
matrix, $\tilde{\mathbf{F}}$, is $\sqrt{T}$ times the eigenvectors
corresponding to the $r$ largest eigenvalues of the $T\times T$ matrix $%
\mathbf{Y}\mathbf{Y}^{\prime }$. It is well-known that this solution is not
unique, that is, any orthogonal rotation of $\tilde{\mathbf{F}}$ is also a
solution. See \cite{bai2008large} for more details.

The following theorem is a modified version of Theorem 1 in \cite%
{bates2013consistent} and shows that, under the assumptions previously
stated, it is possible to consistently estimate any rotation of the factors
by principal components even if the loadings are time-varying.

\begin{theorem}
\label{propoCP} Under Assumptions A-E, there exists an $r\times r$ matrix $H$,
such that 
\begin{equation}  \label{propoCP1}
T^{-1}\sum_{t=1}^{T}\|\tilde{\mathbf{F}}_t-H{^{\prime }}\mathbf{F}%
_t\|^2=O_p(R_{NT}),
\end{equation}
as $N,T\rightarrow\infty$, where $R_{NT}=\max\left\{\frac{1}{N}, \frac{1}{%
NT}, \frac{K_{1NT}}{N^2}, \frac{K_{2NT}}{N^2T^2},\frac{K_{3NT}}{N^2T^2}%
\right\}$ with $K_{1NT}$, $K_{2NT}$, and $K_{3NT}$ defined in the Assumption
D. Furthermore,\newline
$$H=(\boldsymbol{\Lambda}_0^{^{\prime }}\boldsymbol{\Lambda}_0/N)(\mathbf{F}%
^{\prime }\tilde{\mathbf{F}}/T)V_{NT}^{-1},$$ where $V_{NT}$ is a diagonal
matrix of the $r$ largest eigenvalues of the matrix $(NT)^{-1}\mathbf{Y}%
\mathbf{Y}^{\prime }$.
\end{theorem}
\noindent\textbf{Proof.} See the appendix \ref{AP1}.

Theorem \ref{propoCP} points out that the average squared deviation between
the estimated factors and the space spanned by a rotation of the actual
factors will vanish at rate $R_{NT}$, which is similar to that in \cite%
{bai2002determining}. Note that from (\ref{propoCP1}), the estimated common
factors, $\tilde{\mathbf{F}}_t$, are identified through a rotation, then,
principal components converge to a rotation of the actual common factors $H{%
^{\prime }}\mathbf{F}_t$.

\subsection{The nonstationary model} \label{subsec:nonstationary}

In many areas, as in economics and finances, finding strong evidence of
nonstationary processes has repeatedly been reported in many empirical studies. Consequently, it is natural to think that many panel data may include
nonstationary economic or financial variables. There has been some debate
concerning the use of differenced variables. The main argument being
discussed is whether differencing the series causes a severe loss of
information. 

In this section, we allow for nonstationary variables in model \ref{3Mod}. Then, assume that $Y_t$ is $I(d)$. Note that while assumptions B-E do not differ under the
nonstationary setup, we should modify the factor structure's assumption. With this in mind, we follow the assumption 1 in \cite{pena2006nonstationary} to define the $r_1$ nonstationary common factors as follows

\begin{myassump}{F} \textbf{Nonstationary Factors:}\vspace{0.5cm}
	 
$	\begin{array}{l}
		(1-L)^{d} \mathbf{f}_{1, t}=\mathbf{\mu}+\mathbf{u}_{t}, \\
		\mathbf{u}_{t}=\Psi(L) \mathbf{a}_{1, t},
		\end{array}	$\vspace{0.5cm}
		
		\noindent where $L$ is the lag operator, $d$ is a positive integer, $\mathbf{\mu}$ is a $r_{1} \times 1$ vector of drifts, $\mathrm{E}\left(\mathbf{a}_{1, t}\right)=\mathbf{0}, \operatorname{var}\left(\mathbf{a}_{1, t}\right)=$ $\Sigma_{1}>0, \mathrm{f}_{1,-(d-1)}=\mathrm{f}_{1,-(d-2)}=\cdots=\mathrm{f}_{1,0}=0, \sum i\left\|\Psi_{i}\right\|<\infty$ and $\|\mathbf{M}\|=\left[\operatorname{tr}\left(\mathbf{M}^{\prime} \mathbf{M}\right)\right]^{1 / 2}$ for
		any matrix or vector M. Define $\Psi(1)=\sum_{i=0}^{\infty} \Psi_{i}$ with\\ $\operatorname{rank}(\Psi(1))=r_{1}.$
\end{myassump}

To estimate the nonstationary model, we use the methodology proposed by \cite{pena2006nonstationary}, who use generalized covariance matrices defined as 
\begin{equation}
\mathbf{C}_{y}(k)=\frac{1}{T^{2d+d^{\prime }}}\sum_{t=k+1}^{T}(\mathbf{Y}%
_{t-k}-\bar{\mathbf{Y}}_{t})(\mathbf{Y}_{t}-\bar{\mathbf{Y}}_{t}),  \label{VG}
\end{equation}%
where $\bar{\mathbf{Y}}=\frac{1}{T}\sum_{t=1}^{T}{\mathbf{Y}}_{t}$ and $%
d^{\prime }$ can be either 0 or 1. 

Under this framework, consistent factor estimates are given by 
\begin{equation}
\hat{\mathbf{F}}=\mathbf{Y}\hat{\boldsymbol{\Lambda }},  \label{FPP}
\end{equation}%
where $\hat{\mathbf{F}}=(\hat{\mathbf{F}}_{1},\hat{\mathbf{F}}_{2},\ldots ,%
\hat{\mathbf{F}}_{T})^{\prime }$ is a $T\times r$ matrix, $\mathbf{Y}=(%
\mathbf{Y}_{1},\ldots ,\mathbf{Y}_{T})^{\prime }$ is a $T\times N$ matrix
and $\hat{\boldsymbol{\Lambda }}$ is a $N\times r$ matrix composed by the
first $r$ eigenvectors of $\mathbf{C}_{y}(k)$. Following the same reasoning
as in Theorem \ref{propoCP}, it can be shown that the average squared
deviation between the estimated factors (\ref{FPP}) and space spanned by a
rotation of the actual factors will vanish as $(N,T)\rightarrow \infty .$

\subsection{Wavelets} \label{secon}

The basic idea of a wavelet is to construct infinite collections of
translated and scaled versions of the scaling function $\phi (t)$ and the
wavelet $\psi (t)$ such as $\phi _{jk}(t)=2^{j/2}\phi (2^{j}t-k)$, and $\psi
_{jk}(t)=2^{j/2}\psi (2^{j}t-k)$ for $j,k\in \mathbb{Z}$. Suppose that $%
\{\phi _{lk}(\cdot )\}_{k\in \mathbb{Z}}\cup \{\psi _{jk}(\cdot )\}_{j\geq
l;k\in \mathbb{Z}}$ forms an orthonormal basis of $L^{2}(\mathbb{R})$, for
any coarse scale $l$. A key point is to construct $\phi $ and $\psi $ with a
compact support that generates an orthonormal system, which has location in
time-frequency. From this, we can get parsimonious representations for a
wide class of wavelet functions, further details in \cite{changmore2005}, and \cite%
{portomoreaub2008}. 

In some applications, these functions are
defined in a compact set in $[0,1]$ for
functions $\lambda _{ij}(u)$, for $i=1,\ldots ,N$ and $j=1,\ldots ,r$
defined in (\ref{3Mod2}). Then, it is necessary to consider an orthonormal
system that generates $L^{2}[0,1]$. For the construction of these
orthonormal systems, we follow the procedure by  \cite{cohen1995wavelets},
that  generates multiresolution levels $\tilde{V}_{0}\subset \tilde{V}%
_{1}\subset \cdots $, where the spaces $\tilde{V}_{j}$ are generated by $%
\tilde{\psi}_{jk}.$ Negative values of $j$ are not necessary since $\tilde{%
\phi}=\tilde{\phi}_{00}=1$, and if $j\leq 0$, $\tilde{\psi}_{jk}(u)=2^{-j/2}$%
, see \cite{vidakovic2009statistical} for more details and different
approaches. Therefore, for any function $\lambda (u)\in L^{2}[0,1]$, can be expand in series of orthogonal functions 
\begin{equation}
\lambda (u)=\alpha _{00}\phi (u)+\sum_{j\geq 0}\sum_{k\in I_{j}}\beta
_{jk}\psi _{jk}(u),  \label{EQ3}
\end{equation}%
where we take $l=0$ and $I_{j}=\{k:k=0,\ldots ,2^{j}-1\}$. For each $j$, the
set $I_{j}$ generates values of $k$ such that $\beta _{jk}$ belongs to the
scale $2^{j}$. For example, for $j=3$, there are eight wavelet coefficients
in the scale $2^{3}$, whereas for $j=2$, only four coefficients in the scale 
$2^{2}$.

Some applications consider the equation in (\ref{EQ3}) for a maximum
resolution level $J$, through 
\begin{equation}
\lambda (u)\approx \alpha _{00}\phi (u)+\sum_{j=0}^{J-1}\sum_{k\in
I_{j}}\beta _{jk}\psi _{jk}(u).  \label{EQ5}
\end{equation}

In this way, the function $\lambda (u)$ approximates to the space $\tilde{V}%
_{J}$. We use ordinary wavelets as in \cite%
{dahlhaus1997identification} because the performance is suitable in the
case of smooth functions. Particularly, we employ Daubechies ($D8$, hereafter) and Haar
wavelets of compact supports.

\section{Estimation of time-varying loadings by wavelets} \label{S33}
We consider the process in (\ref{3Mod}) with $r$ common factors $(r<N)$ to
discuss the estimation procedure of the time-varying loadings. From now on,
we consider the loading matrix, $[\boldsymbol{\Lambda}(u)+\boldsymbol{%
\Lambda}_0]$, as a unique function over time $\boldsymbol{\Lambda}(u)$,
given by 
\begin{equation}  \label{modos}
\mathbf{Y}_t=\boldsymbol{\Lambda}(u)\mathbf{F}_t+\mathbf{e}_t,
\end{equation}
with $t=1,2,\ldots,T$, and $u=t/T\in[0,1].$ In matrix form, we have
\begin{equation}  \label{MATRIX}
\left[%
\begin{array}{c}
Y_{1t} \\ 
Y_{2t} \\ 
\vdots \\ 
Y_{rt} \\ 
\vdots \\ 
Y_{Nt} \\ 
\end{array}%
\right] = \left[%
\begin{array}{cccc}
\lambda_{11}(u) & \lambda_{12}(u) & \ldots & \lambda_{1r}(u) \\ 
\lambda_{21}(u) & \lambda_{22}(u) & \ldots & \lambda_{2r}(u) \\ 
\vdots & \vdots & \ddots & \vdots \\ 
\lambda_{r1}(u) & \lambda_{r2}(u) & \ldots & \lambda_{rr}(u) \\ 
\vdots & \vdots & \ddots & \vdots \\ 
\lambda_{N1}(u) & \lambda_{N2}(u) & \ldots & \lambda_{Nr}(u) \\ 
&  &  & 
\end{array}%
\right] \left[%
\begin{array}{c}
F_{1t} \\ 
F_{2t} \\ 
\vdots \\ 
F_{rt} \\ 
\end{array}%
\right]+ \left[%
\begin{array}{c}
e_{1t} \\ 
e_{2t} \\ 
\vdots \\ 
e_{rt} \\ 
\vdots \\ 
e_{Nt} \\ 
\end{array}%
\right],
\end{equation}
where $\mathbf{Y}_t$ is an $N-$dimensional vector of time series, $%
\boldsymbol{\Lambda}(u)$ is the time-varying loading matrix with $%
\lambda_{ij}(u)\in L^2[0,1]$, for $i=1,2,\ldots,N$, and $j=1,2,\ldots,r$. $%
\mathbf{F}_t$ is the common factor, and $\mathbf{e}_t$ is the idisyncratic
process. From (\ref{modos}), and the Assumption E, the structure of the
covariance matrix of the process $\mathbf{Y}_t$ is written as 
\begin{equation*}  \label{VARY}
\Gamma_Y(u)=\boldsymbol{\Lambda}(u)\Gamma_F\boldsymbol{\Lambda}^{\prime
}(u)+\Gamma_e, \ \ \ \ \forall{}u\in[0,1],
\end{equation*}
that means, the variance of the common component is ${\Gamma}_X(u)={%
\boldsymbol{\Lambda}}(u) {\Gamma}_F{\boldsymbol{\Lambda}}^{\prime }(u)$.

For the construction of the time-varying loadings estimates, we first assume
that the estimator of the $r$ common factors are obtained by PCA of the $N-$%
dimensional time series $\mathbf{Y}_t$. Then, functions of time-varying
loadings are approximated in series of orthogonal wavelets as in (\ref{EQ5}%
), for a fixed resolution level $J<T$, 
\begin{equation}  \label{WAV2}
\lambda_{mn}(u)= \alpha_{00}^{(mn)}\phi(u) + \sum_{j=0}^{J-1}\sum_{k\in
I_j}\beta_{jk}^{(mn)}\psi_{jk}(u).
\end{equation}

The values of $j,k$ vary depending on the resolution level in the wavelet
decomposition. We choose the maximum resolution $J$, such that $2^{J-1}\leq 
\sqrt{T} \leq 2^{J}$, see \cite{dahlhaus1997identification} for details of
this selection. In practice, the coefficients $\alpha_{00}^{(mn)},
\beta_{00}^{(mn)}, \beta_{10}^{(mn)}, \ldots, \beta_{J-1,2^J-1}^{(mn)}$ are
obtained for a particular estimation method. In this paper, we use GLS to
estimate these coefficients and to reconstruct the loadings functions.

Let $\mathbf{Y}_t=(Y_{1t}, Y_{2t}, \ldots, Y_{Nt})$ be $N$ time series with $%
t=1,2,\ldots,T$ which are generated by 
\begin{equation}  \label{MODGERAL}
\mathbf{Y}_t=\boldsymbol{\Lambda}(u)\tilde{\mathbf{F}}_t+\mathbf{e}_t,
\end{equation}
where the $r$ common factors, $\tilde{\mathbf{F}}_t$, are estimated by
principal components. Each loading function $\lambda_{mn}(u)$ is written as
in (\ref{WAV2}), then when plugging each $\lambda_{mn}(u)$ into (\ref{MATRIX}%
), we have

\begin{gather}  \label{GERAL}
\underbrace{\left[%
\begin{array}{c}
{Y}_{11} \\ 
\vdots \\ 
{Y}_{1T} \\ 
\vdots \\ 
{Y}_{r1} \\ 
\vdots \\ 
{Y}_{rT} \\ 
\vdots \\ 
{Y}_{N1} \\ 
\vdots \\ 
{Y}_{NT} \\ 
\end{array}%
\right]}_{vec(\mathbf{Y})} = \underbrace{\left[%
\begin{array}{ccccccccccccc}
\Psi_{\tilde{\mathbf{F}}}^{(1)} & \Psi_{\tilde{\mathbf{F}}}^{(2)} & \ldots & 
\Psi_{\tilde{\mathbf{F}}}^{(r)} & \ldots & \mathbf{O} & \mathbf{O} & \ldots
& \mathbf{O} & \mathbf{O} & \mathbf{O} & \ldots & \mathbf{O} \\ 
\mathbf{O} & \mathbf{O} & \mathbf{O} & \mathbf{O} & \ldots & \mathbf{O} & 
\mathbf{O} & \ldots & \mathbf{O} & \mathbf{O} & \mathbf{O} & \ldots & 
\mathbf{O} \\ 
\vdots & \vdots & \vdots & \vdots & \ddots & \vdots & \vdots &  & \vdots & 
\vdots & \vdots &  & \vdots \\ 
\mathbf{O} & \mathbf{O} & \mathbf{O} & \mathbf{O} & \ldots & \Psi_{\tilde{%
\mathbf{F}}}^{(1)} & \Psi_{\tilde{\mathbf{F}}}^{(2)} & \ldots & \Psi_{\tilde{%
\mathbf{F}}}^{(r)} & \mathbf{O} & \mathbf{O} & \ldots & \mathbf{O} \\ 
\vdots & \vdots & \vdots & \vdots &  & \vdots & \vdots &  & \vdots & \vdots
& \vdots &  & \vdots \\ 
\mathbf{O} & \mathbf{O} & \mathbf{O} & \mathbf{O} & \ldots & \mathbf{O} & 
\mathbf{O} & \ldots & \mathbf{O} & \Psi_{\tilde{\mathbf{F}}}^{(1)} & \Psi_{%
\tilde{\mathbf{F}}}^{(2)} & \ldots & \Psi_{\tilde{\mathbf{F}}}^{(r)} \\ 
&  &  &  &  &  &  &  &  &  &  &  & 
\end{array}%
\right]}_{\boldsymbol{\Theta}} \underbrace{\left[%
\begin{array}{c}
\boldsymbol{\beta}^{(1)} \\ 
\boldsymbol{\beta}^{(2)} \\ 
\boldsymbol{\beta}^{(3)} \\ 
\vdots \\ 
\boldsymbol{\beta}^{(r)} \\ 
\vdots \\ 
\boldsymbol{\beta}^{(N)} \\ 
\end{array}%
\right]}_{\boldsymbol{\beta}}+ \underbrace{\left[%
\begin{array}{c}
{e}_{11} \\ 
\vdots \\ 
{e}_{1T} \\ 
\vdots \\ 
{e}_{r1} \\ 
\vdots \\ 
{e}_{rT} \\ 
\vdots \\ 
{e}_{N1} \\ 
\vdots \\ 
{e}_{NT} \\ 
\end{array}%
\right],}_{vec(\mathbf{e})}
\end{gather}
where 
\begin{equation*}
\Psi_{\tilde{\mathbf{F}}}^{(i)}=\left[%
\begin{array}{cccc}
\phi(1/T){\tilde{F}}_{i1} & \psi_{00}(1/T){{\tilde{F}}}_{i1} & \ldots & 
\psi_{J-1,2^J-1}(1/T){\tilde{F}}_{i1} \\ 
\phi(2/T){{\tilde{F}}}_{i2} & \psi_{00}(2/T){{\tilde{F}}}_{i2} & \ldots & 
\psi_{J-1,2^J-1}(2/T){{\tilde{F}}}_{i2} \\ 
\vdots & \vdots & \ddots & \vdots \\ 
\phi(T/T){{\tilde{F}}}_{iT} & \psi_{00}(T/T){{\tilde{F}}}_{iT} & \ldots & 
\psi_{J-1,2^J-1}(T/T){{\tilde{F}}}_{iT} \\ 
&  &  & 
\end{array}%
\right],
\end{equation*}
are $T\times 2^J$ matrices for $i=1,2,\ldots,r$ and $\mathbf{O}$ is $T\times
2^J$ null matrix.

Let $\Psi_{r{\tilde{F}}}=[\Psi_{{\tilde{F}}}^{(1)},\ldots,\Psi_{{\tilde{F}}%
}^{(r)}]$ be a $T\times r2^J$ matrix, then 
\begin{equation*}  \label{planeja}
\boldsymbol{\Theta}=\mathbb{I}_N\otimes\Psi_{r{\tilde{F}}}
\end{equation*}
is $NT\times Nr2^J$ matrix that depends on the estimated factors, $\tilde{%
\mathbf{F}}_t$, the wavelets $\psi(u)$, and the resolution level $J$, with
vector of parameters $\boldsymbol{\beta}^{(m)}=\left(\boldsymbol{\beta}%
^{(m1)}, \boldsymbol{\beta}^{(m2)}, \ldots ,\right.$ $\left. \boldsymbol{%
\beta}^{(mr)}\right)^{\prime }$ of dimension $r2^J\times1$ for $%
m=1,2,\ldots,N$, where $\boldsymbol{\beta}^{(mn)}=\left(\alpha_{00}^{(mn)},%
\right.$ $\left.\beta_{00}^{(mn)}, \beta_{10}^{(mn)}, \ldots,
\beta_{J-1,2^J-1}^{(mn)}\right)^{\prime }$.

Each $\boldsymbol{\beta}^{(m)}$ is composed by the wavelets coefficients of
the $m-$th row of the matrix $\boldsymbol{\Lambda}(u)$. Therefore, the total
number of wavelets parameters to be estimated is $2^JNr$.

Hence, the model in (\ref{GERAL}) can be represented in a linear model form
as 
\begin{equation*}  \label{OLSGERALEQ}
vec(\mathbf{Y})=\boldsymbol{\Theta}\boldsymbol{\beta}+vec(\mathbf{e}),
\end{equation*}
where $vec(\mathbf{Y})$ is the response vector and $\boldsymbol{\Theta}$ is
the usual design matrix in regression analysis. Assuming that the covariance
matrix of the idiosyncratic errors, $\Gamma_e$, is known, then the GLS
estimator of the coefficients $\boldsymbol{\beta}$ is given by 
\begin{equation*}  \label{BETA2}
\hat{\boldsymbol{\beta}}=(\boldsymbol{\Theta}^{\prime }\Sigma_e^{-1}%
\boldsymbol{\Theta})^{-1}\boldsymbol{\Theta}^{\prime }\Sigma_e^{-1}vec(%
\mathbf{Y}),
\end{equation*}
where $\Sigma_e$ is a $NT\times NT$ matrix defined as 
\begin{equation*}
\Sigma_e=\Gamma_e\otimes\mathbb{I}_T=\left[%
\begin{array}{cccc}
\mathbb{I}_T\gamma_{e,11} & \mathbb{I}_T\gamma_{e,12} & \ldots & \mathbb{I}%
_T\gamma_{e,1N} \\ 
\mathbb{I}_T\gamma_{e,21} & \mathbb{I}_T\gamma_{e,22} & \ldots & \mathbb{I}%
_T\gamma_{e,2N} \\ 
\vdots & \vdots & \ddots & \vdots \\ 
\mathbb{I}_T\gamma_{e,N1} & \mathbb{I}_T\gamma_{e,N2} & \ldots & \mathbb{I}%
_T\gamma_{e,NN} 
\end{array}%
\right].
\end{equation*}

These results provide linear estimators of wavelet coefficients for the
time-varying loadings assuming that the covariance matrix of the
idiosyncratic error is known. Some procedures, such as maximum likelihood
methods, are not computationally efficient for estimating such a model since
the number of parameters tends to be very large. In this light, we use GLS
to simplify the implementation.

Note that we can use a different basis of wavelet functions $\phi(u)$, and $%
\psi_{jk}(u)$ for each $\lambda_{mn}(u)$. In the simulation section, we use a
similar basis to simplify the exposition.

\subsection{Estimation algorithm}\label{ALGOR}

The estimation algorithm is as follows:

\begin{description}
\item[Step 1.] Use principal components to estimate $\mathbf{F}_t$ as 
\begin{equation*}
\tilde{\mathbf{F}}=\sqrt{T}[v_1,v_2,\ldots,v_r],
\end{equation*}
where $v_i$ is the eigenvector corresponding to the $i-$th largest
eigenvalue, $\lambda_i$ for $i=1,\ldots,r$, of the matrix $(NT)^{-1}\mathbf{Y%
}\mathbf{Y}^{\prime }$. Here, $\mathbf{Y}$ and $\tilde{\mathbf{F}}$ denote $%
T\times N$ and $T\times r$ matrices, respectively. Next, the loading
function matrix, $\boldsymbol{\Lambda}(t)$, is approximated by wavelets 
\begin{equation*}  \label{WAV}
\lambda_{mn}(t)= \alpha_{00}^{(mn)}\phi(t) + \sum_{j=0}^{J-1}\sum_{k\in
I_j}\beta_{jk}^{(mn)}\psi_{jk}(t),
\end{equation*}
with $I_j=\{k : k=0,1,\ldots,2^j-1\}$, $m=1,\ldots,N$, and $n=1,\ldots,r$.
Then, we write the equation (\ref{MODGERAL}) as 
\begin{equation*}  \label{MLWAV}
vec(\mathbf{Y}) = \boldsymbol{\Theta}(\tilde{\mathbf{F}},\psi,\phi)%
\boldsymbol{\beta} + vec(\mathbf{e}),
\end{equation*}
where $vec(\mathbf{Y})$ and $vec(\mathbf{e})$ are $NT\times1$ vectors, and
the dimensions of $\boldsymbol{\Theta}(\tilde{\mathbf{F}}, \psi,\phi)$ and $%
\boldsymbol{\beta}$ are $(NT\times 2^JNr)$ and $(2^JNr\times1)$,
respectively, where $J$ indicates the resolution level chosen in the wavelet
expansions.
\item[Step 2.] Estimate by GLS the wavelet coefficients as 
\begin{equation*}  \label{COEF}
\hat{\boldsymbol{\beta}}=(\boldsymbol{\Theta}^{\prime}\Sigma^{-1}_e\boldsymbol{\Theta%
})^{-1}\boldsymbol{\Theta}^{\prime}\Sigma^{-1}_e\mathbf{Z},
\end{equation*}
where $\boldsymbol{\Theta}(\tilde{\mathbf{F}},\psi,\phi)\equiv\boldsymbol{%
\Theta}$, $\mathbf{Z}=vec(\mathbf{Y})$, and ${\Sigma}_e=\mathbb{I}_{NT}$ is
used as initial value.
\item[Step 3.] Using the estimated coefficient in the Step 2, the loadings
are obtained as 
\begin{equation*}
\hat{\boldsymbol{\Lambda}}(t)^{(0)}= \{\hat{\lambda}_{mn}^{(0)}(t)\}_{m=1,%
\ldots,N}^{n=1,\ldots,r},
\end{equation*}
where 
\begin{equation*}  \label{ESTINDA}
\hat{\lambda}_{mn}^{(0)}(t)=\hat{\alpha}_{00}^{(mn)}\phi(t) +
\sum_{j=0}^{J-1}\sum_{k\in I_j}\hat{\beta}_{jk}^{(mn)}\psi_{jk}(t).
\end{equation*}
\item[Step 4.] With $\hat{\boldsymbol{\Lambda}}(t)^{(0)}$, obtain residuals, 
$\mathbf{Y}_t-\hat{\boldsymbol{\Lambda}}(t)^{(0)}\tilde{\mathbf{F}}_t=\hat{\textbf{e}}%
_t^{(0)}.$ Then, compute 
\begin{equation*}
\hat{\Gamma}_e^{(0)}=\sum_{t=1}^T\hat{\textbf{e}}_t^{(0)}\hat{\textbf{e}}_t^{(0)^{\prime }}/T.
\end{equation*}
\item[Step 5.] Back to step 2 with ${\Sigma}_e=\hat{\Gamma}_e^{(0)}$.
Iterate $n$-times the procedure to obtain the sequences $\{\hat{\boldsymbol{%
\Lambda}}(t)^{(i)}, \hat{\Gamma}_e^{(i)}\}_{i=1,\ldots,n}$. Stop the
iteration when 
\begin{equation*}
\|\hat{\boldsymbol{\Lambda}}(t)^{(i-1)}-\hat{\boldsymbol{\Lambda}}%
(t)^{(i)}\|<\delta,
\end{equation*}
for any small $\delta>0$, where $\|\cdot\|$ denotes the Frobenius norm for $%
t=1,\ldots,T.$
\end{description}

It is possible to improve the efficiency of the common factors by giving the new value of the loading matrix, regressing the series on loadings to obtain a new estimate of the factors, and then iterate the algorithm. However, this procedure is computationally very intensive for large data sets. As shown in the following two sections, it is unnecessary to implement an expensive iterative procedure to get successful results.  Our algorithm is comparable with the methodology proposed by \cite{mikkelsen2018consistent}, which also does not iterate the algorithm. 

\section{Monte Carlo simulation}\label{sec:montecarlo}

We examine the finite-sample properties of the estimation procedure proposed
above using a Monte Carlo study. The model in (\ref{3Mod}) is generated as 
\begin{eqnarray*}
Y_{it} &=&\boldsymbol{\lambda }_{i}^{\prime }(t)\mathbf{F}_{t}+e_{it},\quad
i=1,\ldots ,N\quad \text{and}\quad t=1,\ldots ,T, \\
{F}_{kt}(1-\theta _{k}L) &=&\eta _{kt},\quad k=1,\ldots ,r.\ \ \boldsymbol{%
\eta }_{t}\sim \mathcal{N}_{r}(0,diag\{\beta _{1}^{2},\ldots ,\beta
_{r}^{2}\}), \\
\mathbf{e}_{t} &\sim &\mathcal{N}_{N}(0,\Gamma _{e}),
\end{eqnarray*}%
where $L$ is the lag operator, $|\beta_i|<1$, for $i=1,\ldots,r$, and the matrix $\Gamma _{e}$ is generated by two
structures; i) $\Gamma _{e}=\{\gamma ^{|i-j|}\}_{i,j=1,\ldots ,N}$, 
a Toeplitz matrix, and ii) a diagonal matrix. Furthermore, at time $t$, $%
Y_{it}$ denotes the $i-$th time series, $\boldsymbol{\lambda }_{i}^{\prime
}(t)=(\lambda _{i1}(t),\ldots ,\lambda _{ir}(t))$ is vector of loadings, which are alternately generated by some smooth functions. We discuss a couple of functions used below. $\mathbf{F}_{t}=(F_{1t},\ldots ,F_{rt})^{\prime }$ is the vector of factors,
and $\boldsymbol{\eta }_{t}=(\eta _{1t},\ldots ,\eta _{rt})^{\prime }$ factor errors and $%
\mathbf{e}_{t}=(e_{1t},\ldots ,e_{Nt})^{\prime }$ are vectors of
idiosyncratic terms which are independent to each other.

A referee kind let us know that Gaussian distributions for both error terms in our Monte Carlo experiment can be relaxed by incorporating other types of distributions. The referee points out that assuming homogeneous distributions on $[F_{min}, F_{max}]$ can help the model be more accurate in some empirical applications. As commented by the referee, a possible estimation method can be the wavelet polynomial chaos method. 

Another possibility to explore our model under a non-Gaussian distribution could be under the state-space modeling. As pointed in \cite{durbin2012time}, it is common to assume normality distribution in the innovation in state-space models because the model is estimated by maximizing a Gaussian log-likelihood, which is evaluated by the Kalman filter. In principle,quasi-maximum likelihood methods can be used when actual distributions of the error terms are non-Gaussian. The possibility of non-Gaussian distributions is beyond the scope of the present
paper and is not further explored. \cite{poncela2021factor} provide an excellent survey on factor extraction using Kalman filter.

In our Monte Carlo study, the model is generated with $N\in \{20,30,100\}$
cross-sectional units and $T\in \{512,1024,2048\}$ sample sizes. We consider
for simplicity only two common factors $(r=2)$. Furthermore, three
values for $\theta_k \in \{0,0.5,1\}$ for $k=1,2$, are considered. Two values
for $\Gamma_{e}$; a Toeplitz matrix  $\Gamma _{e}=Toep$ with $\gamma =0,7$
for correlated noise and $\Gamma _{e}=Diag$ for uncorrelated, where 
a uniform
distribution $U(0.5,1.5)$ generates the entries of the diagonal matrix. Note that simulated data are standardized before extracting the principal components. Common factors are estimated by principal components in cases with $\theta_k<1$, and by the procedure of \cite{pena2006nonstationary} in the case with $\theta_k=1$. All simulations are based on 1000 replications of the model.

We rotate the obtained factors to compare proposed
estimations with the actual factors. The optimal rotation $A^{*}$ is
obtained by maximizing\\ $tr[corr(\mathbf{F},\tilde{\mathbf{F}}A)]$. The solution is given by $A^{*}=VU$ where $V$ and $U$ are orthogonal matrices of the decomposition $corr(F,\tilde{F})=USV^{\prime }$. When the
number of $k$ principal components is not equal to the number of factors $r$%
, we rotate the first $l=\min\{k,r\}$ principal components, see \cite%
{eickmeier2015classical}. Both estimated and simulated factors are
re-scaled to keep the same standard deviation, then 
\begin{equation}  \label{ROTT}
\tilde{\textbf{F}}_k^{*}=\frac{\sigma(F_k)}{\sigma(\tilde{F}_k)}\tilde{\textbf{F}}_k, \ \
k=1,\ldots,r,
\end{equation}
where $\tilde{\textbf{F}}_k$ is the $k-$th column of the matrix of the rotated
principal components $\tilde{\mathbf{F}}A^*$.

As explained before, these rotated factors are now treated as observed
variables in the regression model 
\begin{equation*}
vec(\mathbf{Y})=\boldsymbol{\Theta}(\tilde{\mathbf{F}}^{*},\psi,\phi)%
\boldsymbol{\beta}+vec(\mathbf{e}),
\end{equation*}
to estimate the wavelet coefficients, $\boldsymbol{\beta}$, where $\tilde{%
\mathbf{F}}^{*}=(\tilde{\textbf{F}}_1^{*},\ldots,\tilde{\textbf{F}}_r^{*})$.

To investigate the performance of the estimation procedure, estimated
and simulated factors are compared as follows:
\begin{enumerate}
\item[i)] The precision of the estimation factors is measured by the 
$R_{\tilde{F},F}^{2}$ statistics as in \cite{bates2013consistent},
given by 
\begin{equation}
R_{\tilde{F},F}^{2}=\frac{tr[\mathbf{F}^{\prime }\tilde{\mathbf{F}}(\tilde{%
\mathbf{F}}^{\prime }\tilde{\mathbf{F}})^{-1}\tilde{\mathbf{F}}^{\prime }%
\mathbf{F}]}{tr[\mathbf{F}^{\prime }{\mathbf{F}}]}, \label{TFAC}
\end{equation}%
where $\tilde{\mathbf{F}}$ is the $T\times r$ matrix of estimated factors as
in (\ref{ROTT}) and $\mathbf{F}$ is the $T\times r$ matrix of the actual
factors, that is the simulated ones. This statistics is a multivariate $R^{2}
$ in a regression of the actual factors on the principal components. When
the canonical correlation of the actual and estimated factors tends to one,
then $R_{\tilde{F},F}^{2}\rightarrow 1$ as well.
\item[ii)] We measure the precision of loadings estimates by 
mean square errors (MSE) between estimated and actual loadings, as in 
\cite{motta2011locally}. The MSE is computed as follows 
\begin{equation*}
MSE(v)=(NT)^{-1}\sum_{t=1}^{T}\Vert \hat{\boldsymbol{\Lambda }}^{(v)}(t)-%
\boldsymbol{\Lambda }(t)\Vert ,
\end{equation*}%
for $v=1,\ldots ,1000$. The estimator of the factor loadings matrix, $%
\boldsymbol{\Lambda }(t)$, is chosen by a path such that 
\small{
\begin{equation}
\hat{\boldsymbol{\Lambda }}(t)=\{\hat{\boldsymbol{\Lambda }}%
^{(m)}(t):MSE_{m}=median\{MSE(1),\ldots ,MSE(1000)\}\}.  \label{EQMCAR}
\end{equation}}
\end{enumerate}

\begin{sidewaystable}[htbp]
	\caption{$T\in\{512,1024,2048\}$, $N\in\{20,30,100\}$, and $r=2$. The measure of the consistency of the estimated unobservable factors and the estimated factor loadings are presented in the report.}
	\resizebox{0.9\textwidth}{!}{\begin{minipage}{\textwidth}
			\begin{turn}{0}
				\begin{tabular}{|ccc|cccccc|cccccc|}
					\cmidrule{4-15}    \multicolumn{1}{r}{} &       &       & \multicolumn{12}{c|}{Wavelet functions} \\
					\cmidrule{4-15}    \multicolumn{1}{r}{} &       &       & \multicolumn{6}{c|}{Haar}     & \multicolumn{6}{c|}{D8} \\
					\cmidrule{4-15}    \multicolumn{1}{c}{} &       &       & \multicolumn{2}{c}{$\theta_k = 0$} & \multicolumn{2}{c}{$\theta_k = 0.5$} & \multicolumn{2}{c|}{$\theta_k = 1$}& \multicolumn{2}{c}{$\theta_k = 0$} & \multicolumn{2}{c}{$\theta_k = 0.5$} & \multicolumn{2}{c|}{$\theta_k = 1$} \\
					\cmidrule{1-15}    $N$     & $T$     & $\Gamma_e$ & $R^2_{\tilde{F},F}$    & $MSE_m$   & $R^2_{\tilde{F},F}$    & $MSE_m$ & $R^2_{\tilde{F},F}$    & $MSE_m$  & $R^2_{\tilde{F},F}$    & $MSE_m$   & $R^2_{\tilde{F},F}$    & $MSE_m$ & $R^2_{\tilde{F},F}$    & $MSE_m$ \\
					\midrule
					\multirow{3}[1]{*}{20} 
					& 512   & \multirow{3}[1]{*}{$Diag$} & 0.9341 & 0.0103 & 0.8042 & 0.1970 & 0.7450 & 0.1562 & 0.9047 & 0.0111 & 0.8655 & 0.0023 & 0.7314 & 0.1191 \\
					& 1024  &       & 0.9421 & 0.0076 & 0.8135 & 0.0900 & 0.7501 & 0.1202 &0.9347 & 0.0082 & 0.8764 & 0.0011 & 0.7439 & 0.0876 \\
					& 2048  &       & 0.9432 & 0.0018 & 0.8237 & 0.0052 & 0.7840 & 0.0711 & 0.9470 & 0.0005 & 0.8855 & 0.0009 & 0.8046 & 0.0045 \\
					&           &       &             &           &             &             &           &             &             &  \\
					\multirow{3}[0]{*}{30} 
					& 512   & \multirow{3}[0]{*}{$Diag$} & 0.9236 & 0.0098 & 0.8056 & 0.0128 & 0.7091 & 0.1091 & 0.9460 & 0.0071 & 0.8521 & 0.0127 & 0.7931 & 0.1093 \\
					& 1024  &       & 0.9547 & 0.0051 & 0.8125 & 0.0090 & 0.7112 & 0.0859 & 0.9487 & 0.0047 & 0.8671 & 0.0079 & 0.7963 & 0.0759 \\
					& 2048  &       & 0.9723 & 0.0015 & 0.8268 & 0.0027 & 0.7324 & 0.0738 & 0.9498 & 0.0020 & 0.8691 & 0.0059 & 0.8213 & 0.0028 \\
					&           &       &             &             &             &             &           &             &             &  \\
					\multirow{3}[0]{*}{100} 
					& 512   & \multirow{3}[0]{*}{$Diag$} & 0.9062 & 0.0088 & 0.8019 & 0.0147 & 0.7014 & 0.1198 & 0.9643 & 0.0062 & 0.8674 & 0.0102 & 0.8984 & 0.0993 \\
					& 1024  &                                          & 0.9100 & 0.0042 & 0.8418 & 0.0081 & 0.7020 & 0.0781 & 0.9701 & 0.0037 & 0.9044 & 0.0081 & 0.9105 & 0.0651 \\
					& 2048  &                                          & 0.9241 & 0.0017 & 0.8857 & 0.0035 & 0.7153 & 0.0719 & 0.9812 & 0.0025 & 0.9202 & 0.0054 & 0.9300 & 0.0198 \\
					&           &       &             &             &             &             &           &             &             &  \\
					\multirow{3}[0]{*}{20} 
					& 512   & \multirow{3}[0]{*}{$Toep$} & 0.8631 & 0.0672 & 0.6932 & 0.0891 & 0.6711 & 0.2201 & 0.8911 & 0.0128 & 0.6871 & 0.0593 & 0.6242 & 0.1223 \\
					& 1024  &       & 0.8634 & 0.0501 & 0.6953 & 0.0702 & 0.7012 & 0.1901 & 0.8926 & 0.0100 & 0.7001 & 0.0337 & 0.6832 & 0.0693 \\
					& 2048  &       & 0.8911 & 0.0137 & 0.7723 & 0.0433 & 0.7321 & 0.1642 & 0.8971 & 0.0091 & 0.7307 & 0.0108 & 0.6941 & 0.0031 \\
					&           &       &             &            &              &             &             &             &             &  \\
					\multirow{3}[1]{*}{30} 
					& 512   & \multirow{3}[1]{*}{$Toep$} & 0.8531 & 0.0472 & 0.7984 & 0.0621 & 0.6812 & 0.0912 & 0.9021 & 0.0117 & 0.6815 & 0.0311 & 0.7483 & 0.1114 \\
					& 1024  &       & 0.8671 & 0.0231 & 0.7730 & 0.0539 & 0.7122 & 0.0734 & 0.9126 & 0.0090 & 0.7270 & 0.0276 & 0.7126 & 0.0554 \\
					& 2048  &       & 0.8711 & 0.0092 & 0.8200  & 0.0311 & 0.7212 & 0.0819 & 0.9232 & 0.0068 & 0.7305 & 0.0109 & 0.7531 & 0.0037 \\
					&           &       &             &            &              &             &             &             &             &  \\
					\multirow{3}[1]{*}{100} 
					& 512   & \multirow{3}[1]{*}{$Toep$} & 0.8503 & 0.0398 & 0.8081 & 0.0510 & 0.7294 & 0.0831 & 0.9368 & 0.0101 & 0.7902 & 0.0263 & 0.7566 & 0.0911 \\
					& 1024  &                                          & 0.8602 & 0.0288 & 0.7942 & 0.0495 & 0.7413 & 0.0698 & 0.9404 & 0.0087 & 0.8012 & 0.0201 & 0.7774 & 0.0283 \\
					& 2048  &                                          & 0.8893 & 0.0104 & 0.8491 & 0.0209 & 0.7629 & 0.0503 & 0.9499 & 0.0070 & 0.8125 & 0.0117 & 0.7849 & 0.0325 \\
					\bottomrule
				\end{tabular}
			\end{turn}
	\end{minipage}}\newline
	\justify
	{\scriptsize \textit{Notes:} The DGP is $		Y_{it}=\boldsymbol{\lambda}'_i(t)\textbf{F}_t+e_{it}$, where $i\in \{20,30,100\}$ and $T\in\{512,1024,2048\}$,
		${F}_{kt}(1-\theta_k B)=\eta_{kt}$, with $k\in\{1,2\}$. Idiosyncratic terms are independently generated as $ \boldsymbol{\eta}_{t} \sim\mathcal{N}_1(0, diag\{\beta_1^2, \beta_2^2\})$ with $|\beta_i|<1$, for $i=1,2$ and $\theta_k \in \{0,0.5,1\}$ defining the degree of serial correlation among factors, and $\textbf{e}_t\sim\mathcal{N}_N(0,\Gamma_e)$ with $\Gamma_e$ defined as diagonal and Toeplitz matrices. Common factors are estimated by principal components in cases with $\theta_k<1$, and by the procedure of \cite{pena2006nonstationary} in the case with $\theta_k=1$. $R^2_{\tilde{F},F}$ is the $R^2$ of a regression of actual on estimates factors. $MSE_m$ is the median of the MSE between the actual and estimated factor loadings. Haar and D8 wavelet functions are used in the study. All experiments are based on 1000 replications.\\} 
	\label{EQMC} 
\end{sidewaystable}

Table \ref{EQMC} shows the results of the estimations of (\ref{TFAC}) and (\ref{EQMCAR}). As can be seen, the methodology proposed in this paper performs very well in relatively small samples regardless of size distortion between $N$ and $T$. As seen in Table \ref{EQMC}, $R^2_{\tilde{F},F}$ is relatively high, indicating the satisfactory performance of the estimator. However, precision is a bit reduced when increasing the value of $\theta$. These findings are maintained for both types of wavelets used and even when we allow for cross-correlation between idiosyncratic errors. Furthermore, inspecting the MSE in Table \ref{EQMC}, we find that the MSEs decrease as $T$ increases in all cases, even if the common factors are serially correlated or if idiosyncratic errors are cross-correlated. Furthermore, another finding indicates that, in general, factor loadings using the wavelet D8 perform better than the wavelet Haar. We think that such results are reasonable due to the smoothness of the wavelet D8 in contrast with the other one. The wavelet Haar should be implemented when factor loadings' dynamic have breaks or perhaps some aggressive jumps. We do not go any further in this line. Wavelet Haar's features are part of an another
research and are out of the present scope.

Furthermore, in Figures \ref{loadingshaar} and \ref{loadingsd8}, we display the methodology's performance to estimate the factor loadings. We choose a couple of different smooth functions for each value of $\theta$ and in both types of wavelets for comparison purposes. We consider the following functions: i) $\lambda_{1,12}(t)=0.4\cos {-3\pi t}$, and ii) $\lambda_{2,8}(t)=0.6(0.7\sqrt{t}-0.5\sin {1.2\pi t})$, where $\lambda_{1,12}$ indicates the loading of the first factor of the cross-sectional unit $i=12$, and $\lambda_{2,8}$, the loadings of the second factor of unit $i=8$. Figures display the actual and estimated time-varying loadings and their bootstrap confidence interval at 95\% with $B=100$ replications, following \cite{moura2012}. In such figures, we can see that the methodology works well independently of the value taken in $\theta_k$.

Finally, as a complement of this simulation study, we compare our methodology with the approach of \cite{mikkelsen2018consistent}. Their estimation methodology is similar to ours in the first stage. The difference is in the second stage because instead of using wavelet functions, as in this paper, they employ Kalman filter procedures to estimate the likelihood function. It is worth mentioning that the main difference between both setups lies in the performance of loading factors. While we assume smooth variations in this paper, their approach consists of stationary VAR processes. 

We use the same DGP as before, but we focus only on the following simplest case.
\begin{eqnarray*}
	Y_{it} &=&\boldsymbol{\lambda }_{i}^{\prime }(t)\mathbf{F}_{t}+e_{it},\quad
	i=1,\ldots ,20\quad \text{and}\quad t=1,\ldots ,512, \\
	{F}_{t} &=&\eta _{t},\quad \textrm{where}\;
		\eta_{t}\sim \mathcal{N}(0,1), \\
	\mathbf{e}_{t} &\sim &\mathcal{N}_{N}(0,\Gamma _{e}),
\end{eqnarray*}%
where the matrix $\Gamma _{e}$ is a diagonal matrix, and idiosyncratic terms are independent of each other. As before, the vector of loadings is generated alternately by some smooth functions. We split them into two groups; i) smooth sine/cosine functions and ii) smooth trending functions (linear, square root, exponential, and logs trends). 

Our analysis finds that the methodology proposed by \cite{mikkelsen2018consistent} does not adjust either smooth sinusoidal functions or smooth trending functions in most cases. The worst performance appears when loadings are trending functions, while it is possible to rescue from time to time some proper estimations when loadings behave as sine/cosine functions. Given our results, we argue that the model proposed in this paper seems to be more suitable when loadings have deterministic trends; in contrast, the methodology proposed by \cite{mikkelsen2018consistent} fails to capture this type of behavior in loadings by treating them as a stationary VAR.

\begin{figure}[H]
	\centering
	\includegraphics[width=1\textwidth]{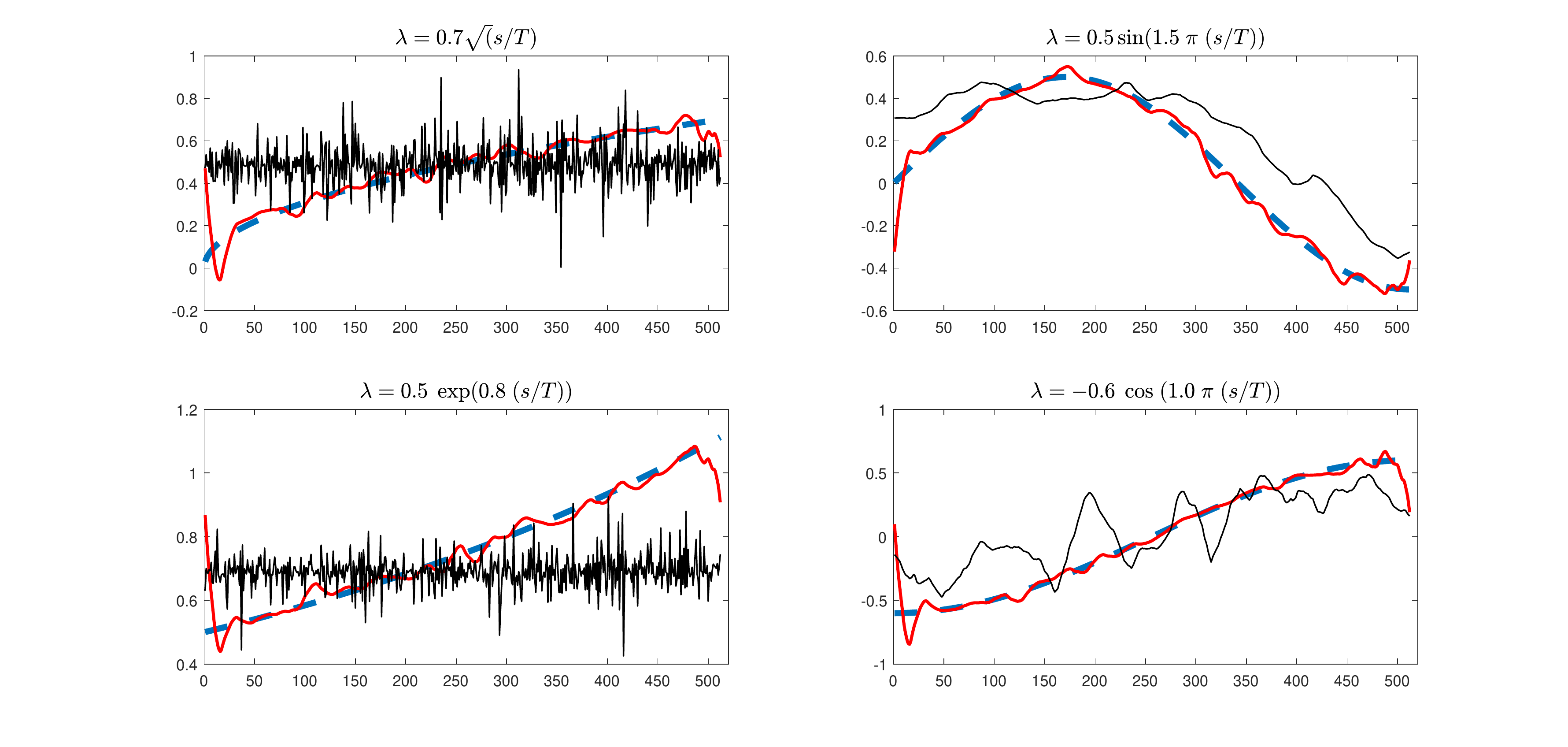}
	\caption{Comparison between the methodology proposed in this paper (solid red line) and that by \cite{mikkelsen2018consistent} (solid black line) to estimate time-varying loadings (dashed blue line) defined by the smooth function written above of each figure.}
	\label{fig:JAKvsDUV}
\end{figure}

We repeated the entire simulation study as in Table \ref{EQMC} using the methodology of \cite{mikkelsen2018consistent}; however, given the inferior performance of loading estimates, we prefer not to augment Table \ref{EQMC} for brevity. Instead, we support this part of the simulation study with Figure \ref{fig:JAKvsDUV} that compares both approaches with four examples of smooth functions, two cases when loadings have trends, and two  with sinusoidal functions. 

\section{Application}

This section applies our methodology to study comovements in loads and prices of the Nord Pool power market.

Nord Pool runs the leading power market in Europe and operates in the
day-ahead and intraday markets. Elspot is the day-ahead auction market, where
participants act in a double auction and submit their supply and demand
orders (spot prices and loads) for each hour of the next day. The
market is in equilibrium when demand and supply curves intersect at the system price and system load for each hour. The hourly system prices
and loads series are announced as 24-dimensional vectors, which are
determined simultaneously.

Electricity markets have particular features that do not exist in any other
commodity market. Remarkably, the non-storability of electricity
provokes that the time series of prices shows excessive volatility, possible
negative prices, and many spikes over time. Other relevant features of electricity prices and loads are the intra-day, week, and year seasonal components, see \cite{weron2007modeling}.

Univariate time series methods have mostly studied hourly electricity prices and loads, see  \cite{Weron20141030} for a rich review. Some
authors have explored these series by multivariate techniques as high dimensional
factor models until the last years.
Using data from the Iberian Electricity Market (MIBEL), seasonal factors have
been extracted in the works of \cite{alonso2011seasonal} and \cite%
{garcia2012forecasting}, whereas \cite{alonso2016electricity} propose to employ model averaging factor models to improve forecasting performance. Pennsylvania - New Jersey - Maryland (PJM) interconnection
market is studied by \cite{maciejowska2015forecasting} who estimate factor
models for forecasting evaluation using hourly and zonal prices.
Furthermore, the Nord Pool power market is studied in the works of \cite{ergemen2016common}, who study the long-term relationship between system prices and loads, and \cite{rodriguez2017estimation}, who use regional prices
in a multi-level setting. 

All these studies assume that factor loadings do not vary over time. In this
sense, we are the first to study the power market with
a factor model with time-varying loadings to the best of our knowledge. Such a setup may help to attract some dynamics not necessarily captured in common factors. We argue that factor loadings
should evolve smoothly, perhaps reacting to the auction market's smooth changes
along the day and along year due to the series' inherent seasonality.  

\begin{figure}
	\centering
	\includegraphics[width=1\textwidth]{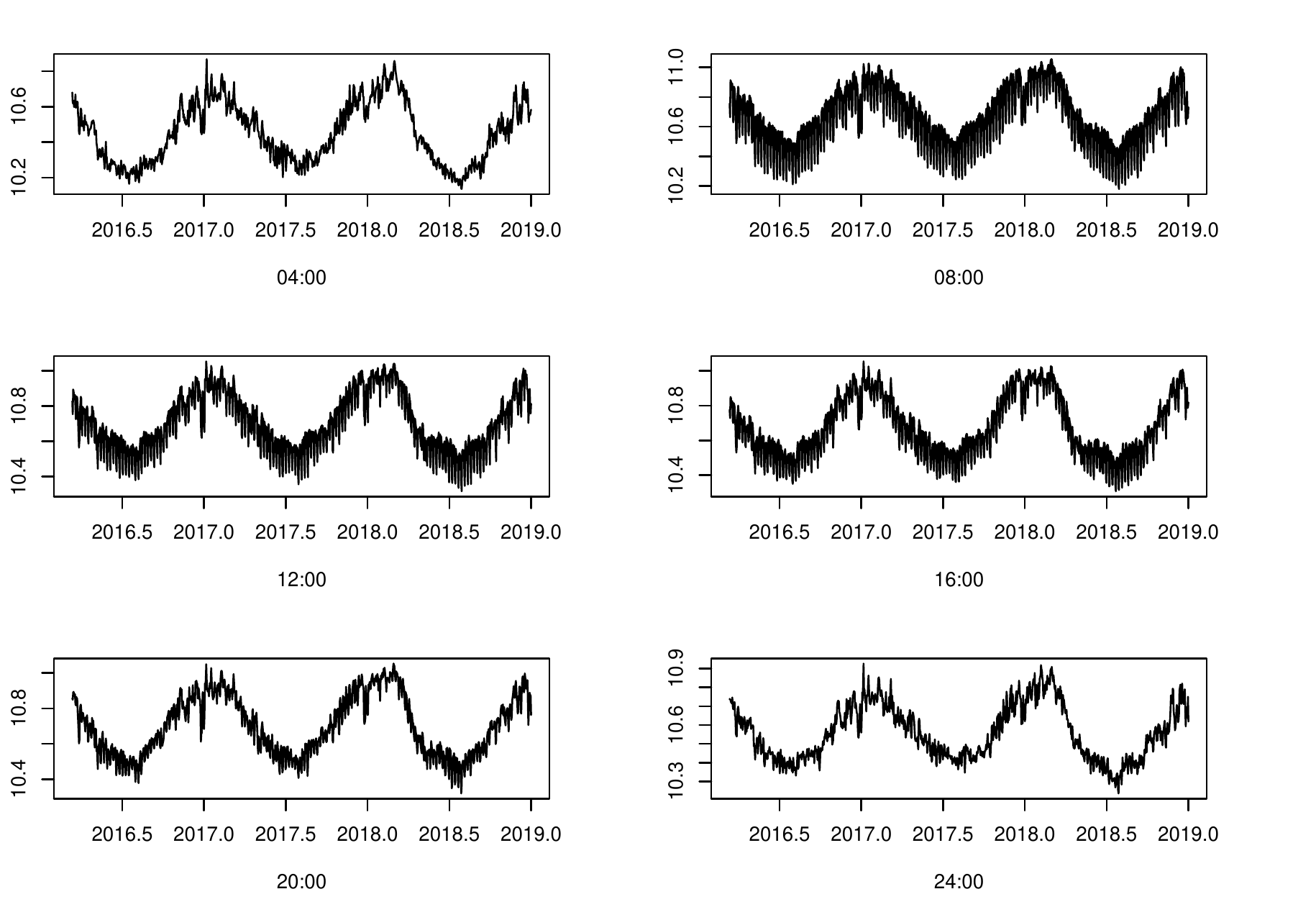}
	\caption{Hourly system loads in logs for six different hours showing working and non-working hours performances, 12 March 2016 to 31 December 2018.}
	\label{fig:loads}
%
	\includegraphics[width=1\textwidth]{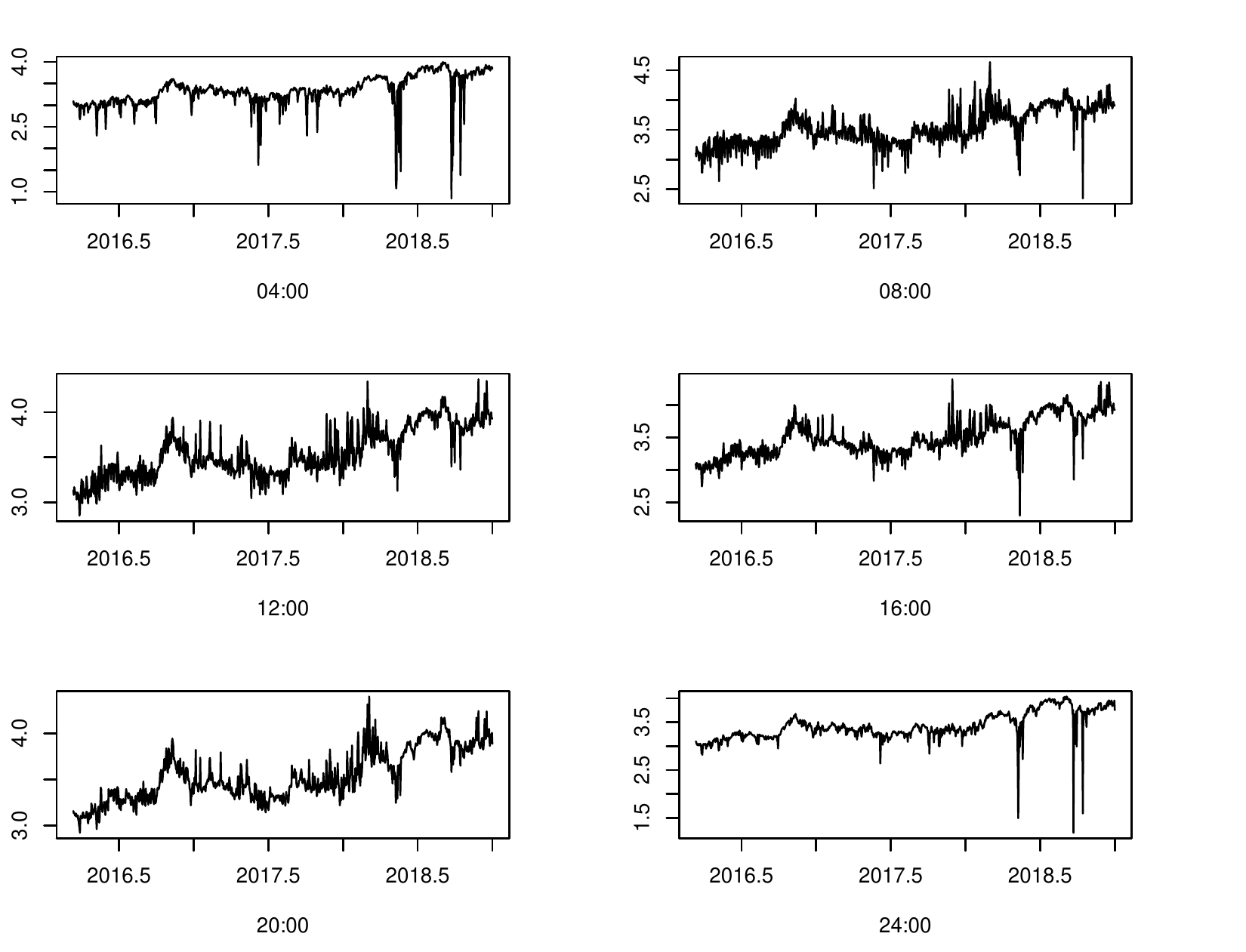}
	\caption{Hourly system prices in logs for six different hours showing working and non-working hours performances, 12 March 2016 to 31 December 2018.}
	\label{fig:prices}
\end{figure}

We consider a balanced panel data set consisting of $N = 24$ hourly prices
and loads for each day from 13th March 2016 to 31st December 2018, yielding
a total of $T = 1024$ daily observations in each hour. We download series from the Nord Pool ftp server, and prices are in Euros
per Mwh of a load. Figures \ref{fig:loads} and \ref{fig:prices} display six
time series in logs from which we can observe some characteristics in both
time series. First, electricity system prices and loads vary differently
along months with a common pattern in their evolutions.
Second, the price series show many spikes which are related to the way how the market operates. Third, seasonal variations
are much more marked in loads than in price series. Fourth, electricity prices
and loads seem to have nonstationary behaviors, then, we set our estimation on the
nonstationary approach discussed in section \ref{subsec:nonstationary}.

We use the
procedure proposed by \cite{alessi2010improved} with first-differenced variables to estimate the number of factors. This methodology introduces a 
multiplicative tuning constant in the penalty function to improve the criteria of 
\cite{bai2002determining}. In line with the literature, we find two common factors in prices, and the same number in 
loads. These factors explain 95\% of the variation in the panel of prices, and 97\% of
loads. These percentages are slightly higher than those found in 
\cite{ergemen2016common}, but our period is shorter and does not cover periods of infrastructure delineation in the Nord Pool power market. Therefore, we estimate the model in (\ref{MODGERAL}) with two common factors for the panel of loads and the panel of prices.  

\begin{figure}[H]
		\centering
	\includegraphics[width=1\textwidth]{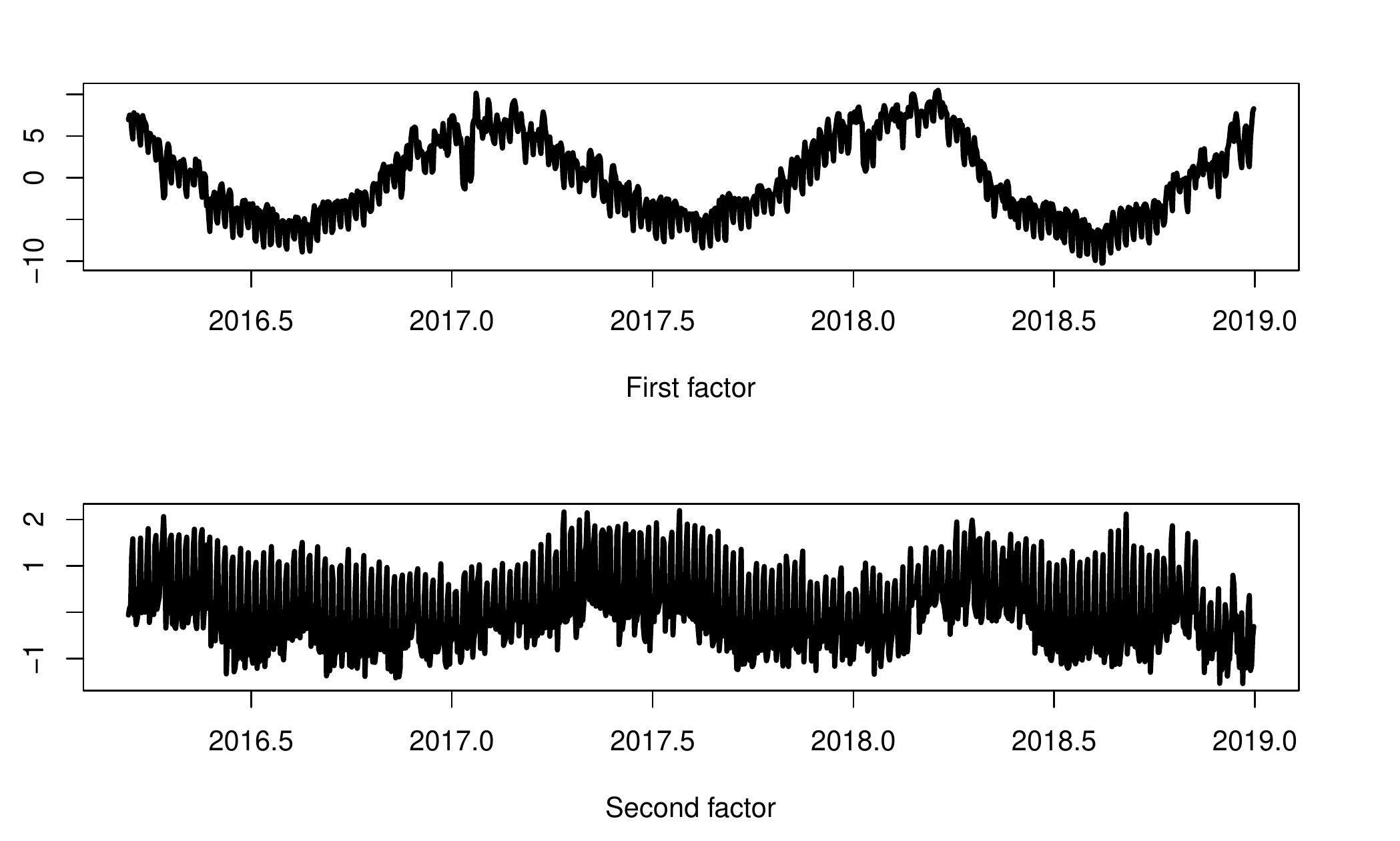}
	\caption{Common factors of hourly system loads.}
	\label{fig:loadsfactors}
\end{figure}

\begin{figure}[H]
	\centering
	\includegraphics[width=1\textwidth]{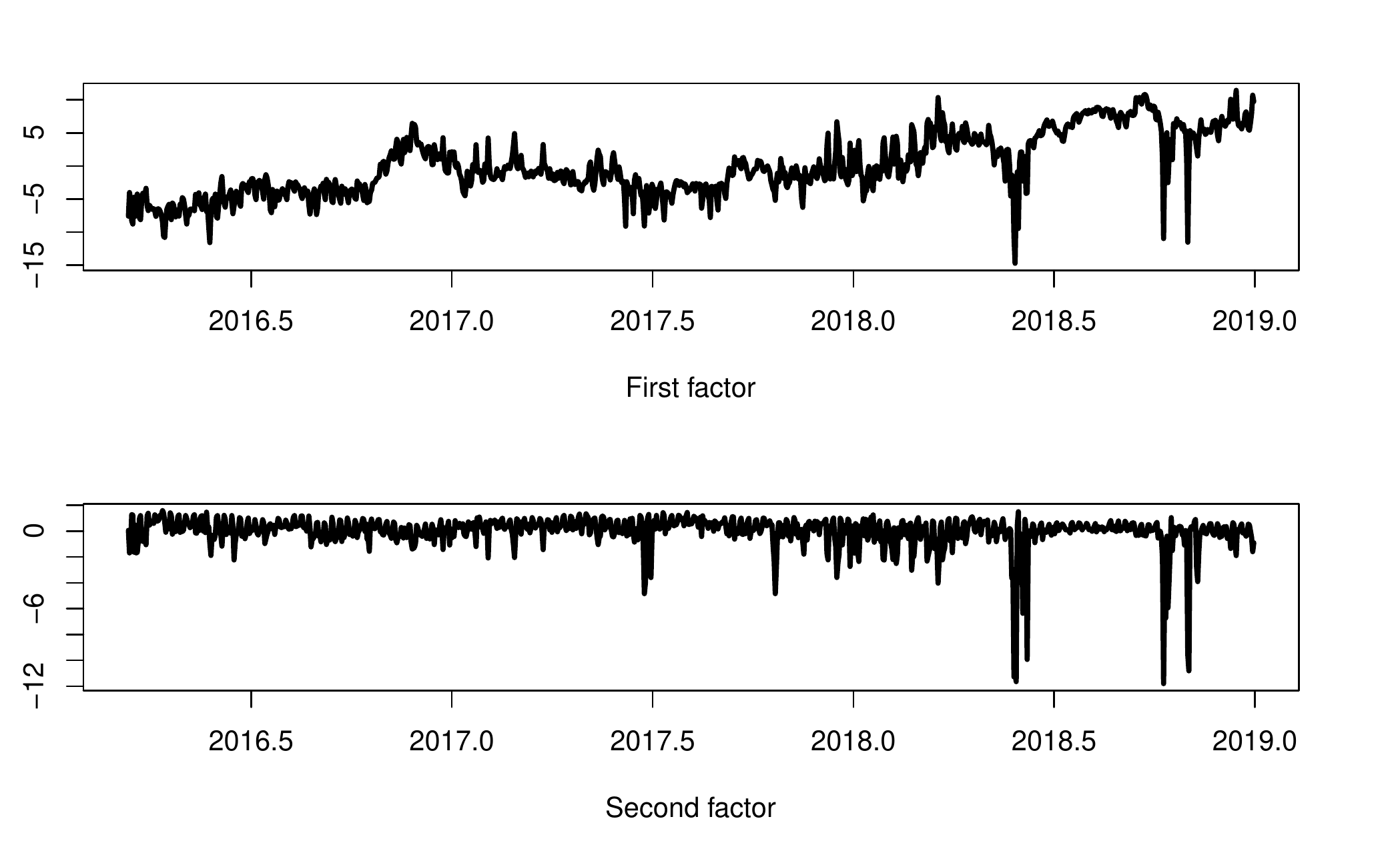}
	\caption{Common factors of hourly system prices.}
	\label{fig:pricesfactors}
\end{figure}

Figures \ref{fig:loadsfactors} and \ref{fig:pricesfactors} display estimates
of common factors of hourly system loads and prices, respectively. As seen in Figure \ref{fig:loadsfactors}, the first factor of hourly loads captures the
strong seasonal component showing possible weekly and monthly periodicities. The second factor seems to capture mainly a more erratic
weekly variability, mostly during working hours, as seen in Figure \ref{fig:loads}. On the other hand, Figure \ref%
{fig:pricesfactors} shows that estimated common factors of hourly system prices exhibit some
stylized facts of the underlying series. Volatility clustering and nonstationary behavior are captured by the first factor, while excessive price spikes in
2018 by March, July, and August are extracted 
mostly by the second factor. \cite{ergemen2016common} also document all these characteristics of common factors.

\begin{figure}[H]
	\centering
	\includegraphics[width=1\textwidth]{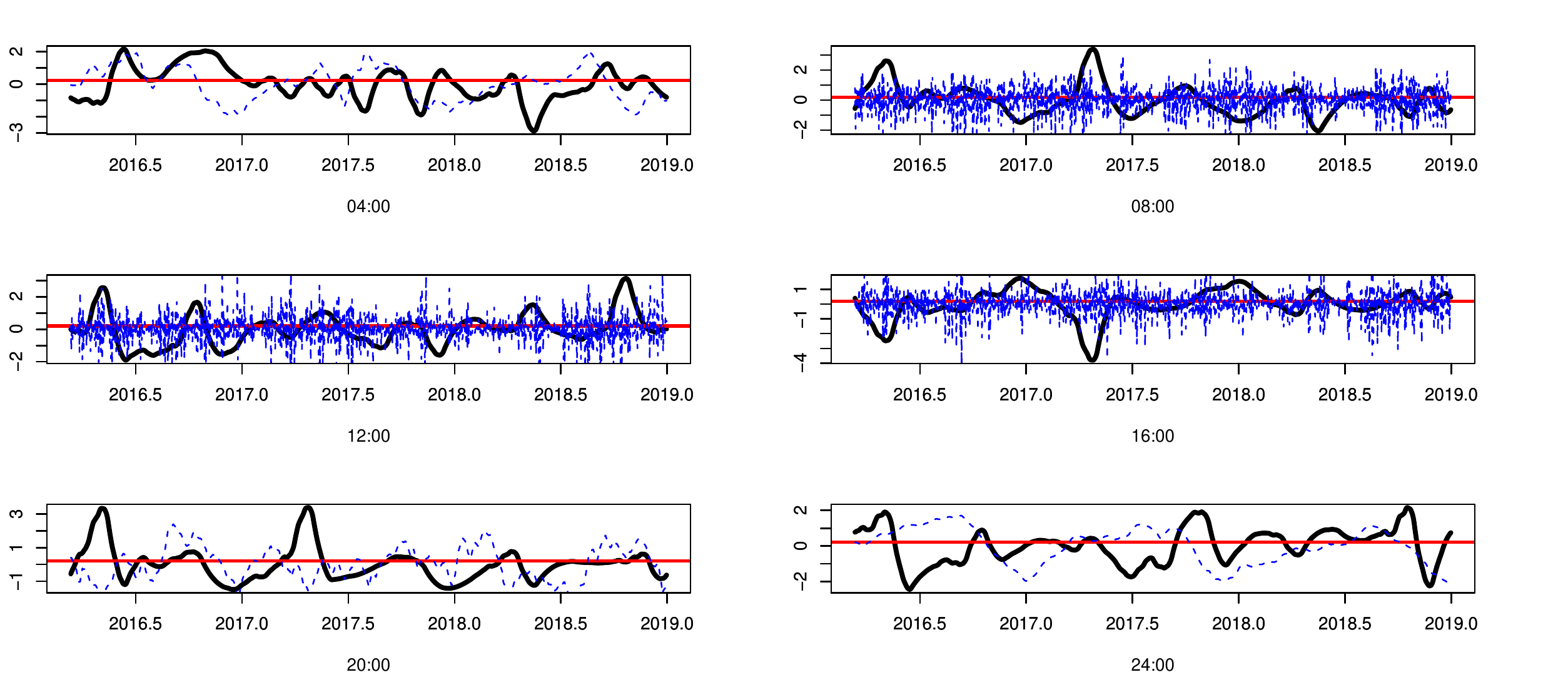}
	\caption{Time-varying loadings of the first factor of hourly system loads. Black solid lines represents loadings estimated by the method proposed in this paper, while blue dot lines represents loadings estimated by \cite{mikkelsen2018consistent}.}
	\label{fig:loadsloading1}
\end{figure}

\begin{figure}
	\centering
	\includegraphics[width=1\textwidth]{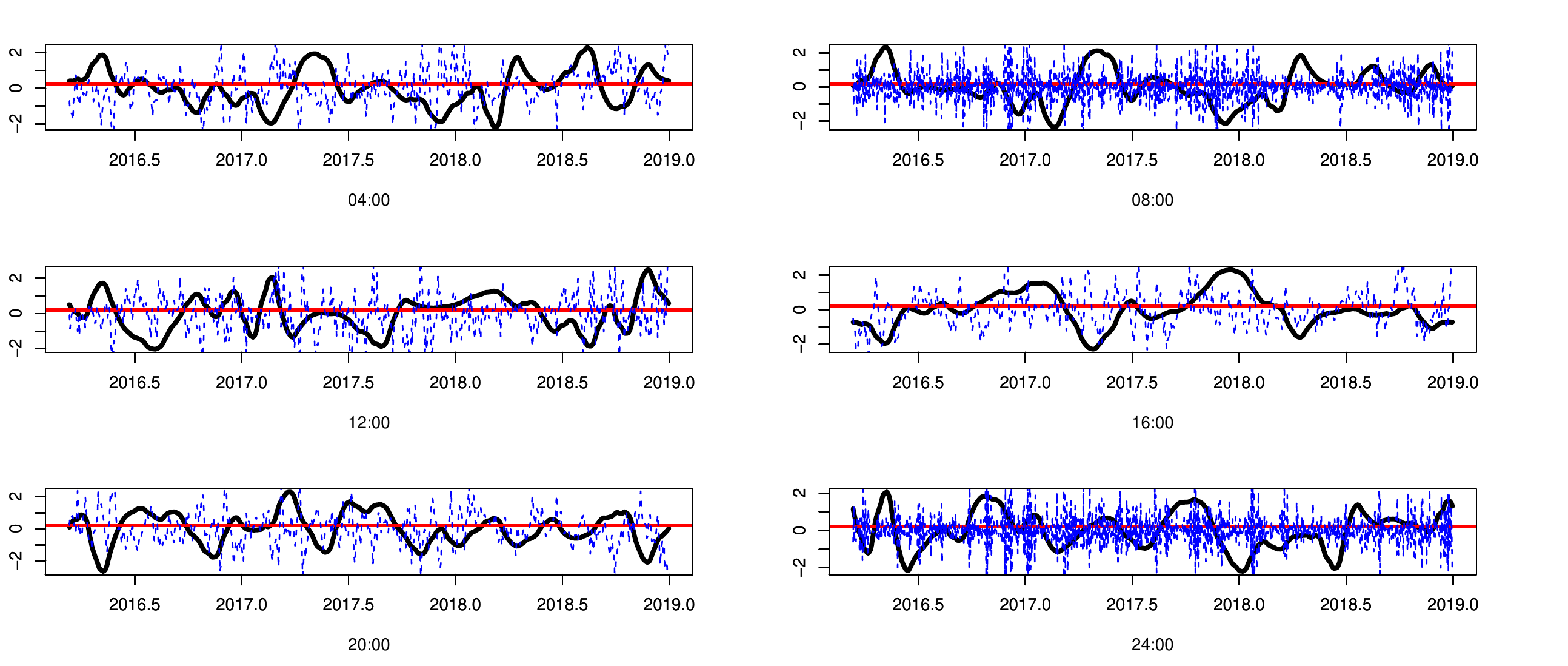}
	\caption{Time-varying loadings of the second factor of hourly system loads. Black solid lines represent loadings estimated by the method proposed in this paper, while blue dot lines represent loadings estimated by \cite{mikkelsen2018consistent}.}
	\label{fig:loadsloading2}
\end{figure}

\begin{figure}
	\centering
	\includegraphics[width=1\textwidth]{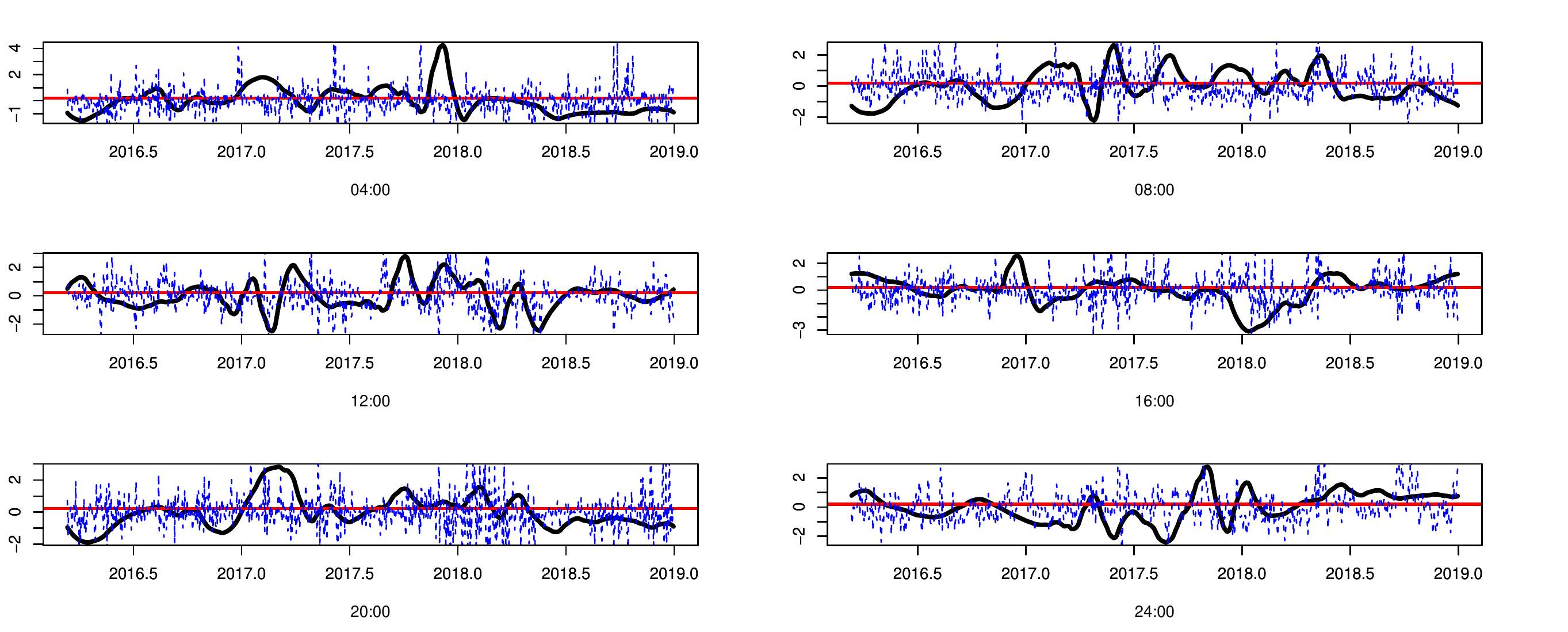}
	\caption{Time-varying loadings of the first factor of hourly system prices. Black solid lines represent loadings estimated by the method proposed in this paper, while blue dot lines represent loadings estimated by \cite{mikkelsen2018consistent}.}
	\label{fig:pricesloading1}
\end{figure}

Figures \ref{fig:loadsloading1} and \ref{fig:loadsloading2} show
time-varying loadings of the first and second common factors of hourly
system loads, respectively. Figures \ref{fig:pricesloading1} and \ref%
{fig:pricesloading2} show the time-varying loadings estimates of hourly
system prices. We only display the results for six hours for exposition purposes: 04:00,  08:00, 12:00, 16:00, 20:00, and 24:00 hrs. These hours help to observe different performances of loadings across working and non-working hours. For comparison purposes, we estimate the model proposed by \cite{mikkelsen2018consistent} using the generalized covariance matrix introduced in (\ref{VG}), in the first stage, to allow for nonstationary variables. We standardized loadings estimates to compare both approaches.

\begin{figure}
	\centering
	\includegraphics[width=1\textwidth]{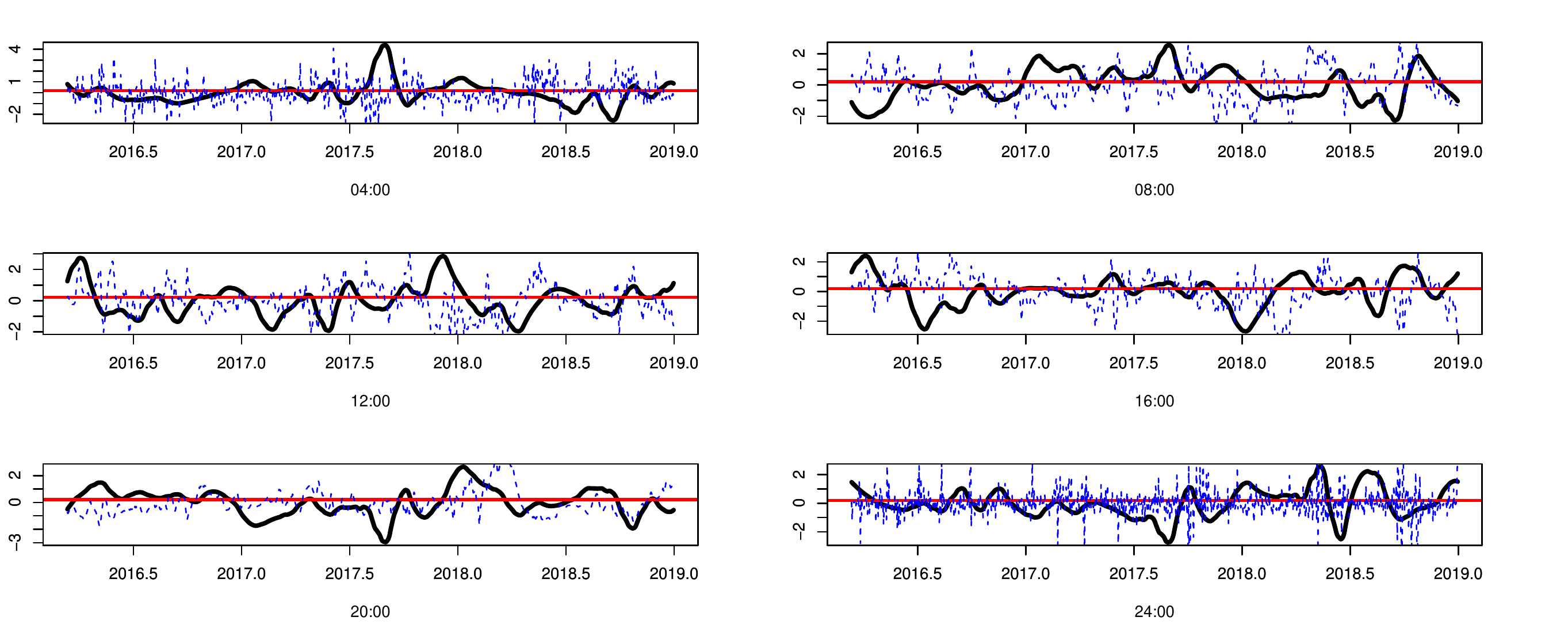}
	\caption{Time-varying loadings of the second factor of hourly system prices. Black solid lines represent loadings estimated by the method proposed in this paper, while blue dot lines represent loadings estimated by \cite{mikkelsen2018consistent}.}
	\label{fig:pricesloading2}
\end{figure}

Further inspection on Figure \ref{fig:loadsloading1} indicates that factor
loadings corresponding to the first common factor of system loads show smooth variations along time when estimating them by the model proposed in this paper (solid black lines in figures). These time-varying loadings capture a kind of readjustment of system loads when the demand reaches a maximum level (in winter) and a minimum level (in summer), which aligns with the market's nature. Moreover, time-varying loadings in the remaining months oscillate smoothly around the mean. In contrast, factor loadings by \cite{mikkelsen2018consistent} (blue dot lines) behave smoothly during non-working hours (see, 04:00, 20:00, and 24:00 hrs.), but the performance is very erratic during working hours (see, 08:00, 12:00, and 16:00 hrs.). Our intuition says that such erratic behavior comes from the volatile behavior of the hourly loads in the same working hours (see again Figure \ref{fig:loads}). Therefore, the main difference between both approaches appears in hours when series behave more volatile. However, it seems intuitive to think that factor loadings should behave smoothly throughout the year without considering the time of day because of changes in the electricity demand. 
 
In turn, factor loadings corresponding to the second common factor of system
loads also have a regular periodic behavior along time, see solid black lines Figure \ref{fig:loadsloading2}. It is interesting to see that loadings behave smoothly independently if the second common factor is much more volatile than the first one (see Figure \ref{fig:loadsfactors}). In contrast, the second approach (blue dot lines in the figure) reacts aggressively along time and could be related to the performance of the second common factor.

Concerning loadings estimates of hourly system prices using the methodology proposed in this paper (solid black lines in Figure \ref{fig:pricesloading1}), at first glance, unlike system loads, we do not find that
factor loadings have precise periodic movements along time. In general, they also behave smoothly as before. It is worth mentioning that some moments where loadings seem to be time-invariant, see, for instance, the last year in Figure \ref{fig:loadsloading1}. Note that this behavior is not captured by loadings estimates using the approach of \cite{mikkelsen2018consistent} (see blue dot lines in the same figures), which provides loadings with a very volatile performance.   

This empirical study leaves many open possibilities for
future research. First, testing if loadings are time-varying or constant over time. This test
would improve the specification of factor models and help to understand the market
behavior. Second, short-term forecasting of prices and loads
is essential for energy markets. Therefore, another research may investigate if
loads/prices forecasts obtained by time-varying factor models are more accurate
than those provided by standard setups. Third, as
discussed in the Monte Carlo study, the factor loadings using the wavelet D8
perform better than Haar in situations where time-series evolve smoothly, as
in the case of system loads. In this sense, a deeper study on electricity
prices could help us understand if the wavelet Haar represents a better
choice for modeling the performance of factor loadings since electricity
prices are more volatile than loads.

To conclude, our setup and methodology can be used in panel data models where unobservable common factors drive the cross-sectional dependence. Recently, \cite{rodriguez2020air} explored this possibility to analyze the dynamic behind air pollution and mobility to disentangle the contagion in the COVID pandemic.  

\section{Concluding remarks}

Factor models' standard approaches assume that factor loadings
are invariant along time. Here, we relaxed such an assumption allowing for time-varying loadings that behave as smooth
and continuous-time functions. The paper is novel because time-varying can capture a more realistic behavior compared to some proposals in the literature that assume that stable stochastic processes drive loadings.  Our estimation methodology is a two-step procedure based on GLS with wavelet functions. We explored stationary and nonstationary setups to the extent of the model's applicability. We recommend using Wavelet D8 estimates in empirical applications with smoothed loadings based on our finite sample analysis. In this sense, Wavelet D8 provides better estimations than Wavelet Haar when factor loadings do not have sudden changes.

Finally, in our empirical study, we focus on the complex dynamics of Nord Pool electricity loads and prices in a large panel of hourly observations. Our findings indicate that factor loadings vary smoothly over time in prices and loads. Notably, in loads, the time-varying loadings have a periodic behavior, which obeys the market's seasonality. The analysis provides relevant insights into the dynamics of the market. Future research will point to how the model proposed can be used from a forecasting perspective. 


%

\clearpage

\bibliographystyle{apa}
\bibliography{biblio}

\begin{thebibliography}{}

\bibitem[\protect\astroncite{Alessi et~al.}{2010}]{alessi2010improved}
Alessi, L., Barigozzi, M., and Capasso, M. (2010).
\newblock Improved penalization for determining the number of factors in
  approximate factor models.
\newblock {\em Statistics \& Probability Letters}, 80(23):1806--1813.

\bibitem[\protect\astroncite{Alonso et~al.}{2016}]{alonso2016electricity}
Alonso, A., Bastos, G., and Garc{\'\i}a-Martos, C. (2016).
\newblock Electricity price forecasting by averaging dynamic factor models.
\newblock {\em Energies}, 9(8):600.

\bibitem[\protect\astroncite{Alonso et~al.}{2011}]{alonso2011seasonal}
Alonso, A.~M., Garc{\'\i}a-Martos, C., Rodr{\'\i}guez, J., and
  Jes{\'u}s~S{\'a}nchez, M. (2011).
\newblock Seasonal dynamic factor analysis and bootstrap inference: application
  to electricity market forecasting.
\newblock {\em Technometrics}, 53(2):137--151.

\bibitem[\protect\astroncite{Bai}{2003}]{bai2003inferential}
Bai, J. (2003).
\newblock Inferential theory for factor models of large dimensions.
\newblock {\em Econometrica}, pages 135--171.

\bibitem[\protect\astroncite{Bai and Han}{2016}]{baihan2016}
Bai, J. and Han, X. (2016).
\newblock Structural changes in high dimensional factor models.
\newblock {\em Frontiers of Economics in China; Beijing}, 11(1):9--39.

\bibitem[\protect\astroncite{Bai and Li}{2016}]{bai2016maximum}
Bai, J. and Li, K. (2016).
\newblock Maximum likelihood estimation and inference for approximate factor
  models of high dimension.
\newblock {\em Review of Economics and Statistics}, 98(2):298--309.

\bibitem[\protect\astroncite{Bai and Ng}{2002}]{bai2002determining}
Bai, J. and Ng, S. (2002).
\newblock Determining the number of factors in approximate factor models.
\newblock {\em Econometrica}, 70(1):191--221.

\bibitem[\protect\astroncite{Bai and Ng}{2004}]{bai2004panic}
Bai, J. and Ng, S. (2004).
\newblock A panic attack on unit roots and cointegration.
\newblock {\em Econometrica}, 72(4):1127--1177.

\bibitem[\protect\astroncite{Bai et~al.}{2008}]{bai2008large}
Bai, J., Ng, S., et~al. (2008).
\newblock Large dimensional factor analysis.
\newblock {\em Foundations and Trends{\textregistered} in Econometrics},
  3:89--163.

\bibitem[\protect\astroncite{Barigozzi et~al.}{2016}]{barigozzi2016non}
Barigozzi, M., Lippi, M., and Luciani, M. (2016).
\newblock Non-stationary dynamic factor models for large datasets.
\newblock {\em arXiv preprint arXiv:1602.02398}.

\bibitem[\protect\astroncite{Bates et~al.}{2013}]{bates2013consistent}
Bates, B.~J., Plagborg-M{\o}ller, M., Stock, J.~H., and Watson, M.~W. (2013).
\newblock Consistent factor estimation in dynamic factor models with structural
  instability.
\newblock {\em Journal of Econometrics}, 177(2):289--304.

\bibitem[\protect\astroncite{Chan et~al.}{2017}]{chan2017factor}
Chan, N.~H., Lu, Y., and Yau, C.~Y. (2017).
\newblock Factor modelling for high-dimensional time series: Inference and
  model selection.
\newblock {\em Journal of Time Series Analysis}, 38(2):285--307.

\bibitem[\protect\astroncite{Chiann and Morettin}{2005}]{changmore2005}
Chiann, C. and Morettin, P.~A. (2005).
\newblock Time-domain estimation of time-varying linear systems.
\newblock {\em Journal of Nonparametric Statistics}, 17(3):365--383.

\bibitem[\protect\astroncite{Cohen and Ryan}{1995}]{cohen1995wavelets}
Cohen, A. and Ryan, R.~D. (1995).
\newblock {\em Wavelets and multiscale signal processing}.
\newblock Springer.

\bibitem[\protect\astroncite{Dahlhaus
  et~al.}{1997a}]{dahlhaus1997identification}
Dahlhaus, R., Eichler, M., and Sandk{\"u}hler, J. (1997a).
\newblock Identification of synaptic connections in neural ensembles by
  graphical models.
\newblock {\em Journal of neuroscience methods}, 77(1):93--107.

\bibitem[\protect\astroncite{Dahlhaus et~al.}{1997b}]{dahlhaus1997fitting}
Dahlhaus, R. et~al. (1997b).
\newblock Fitting time series models to nonstationary processes.
\newblock {\em The annals of Statistics}, 25(1):1--37.

\bibitem[\protect\astroncite{de~A.~Moura et~al.}{2012}]{moura2012}
de~A.~Moura, M.~S., Morettin, P.~A., Toloi, C. M.~C., and Chiann, C. (2012).
\newblock Transfer function models with time-varying coefficients.
\newblock {\em Journal of Probability and Statistics}, 2012:31.

\bibitem[\protect\astroncite{Durbin and Koopman}{2012}]{durbin2012time}
Durbin, J. and Koopman, S.~J. (2012).
\newblock {\em Time series analysis by state space methods}.
\newblock Oxford university press.

\bibitem[\protect\astroncite{Eichler et~al.}{2011}]{eichler2011fitting}
Eichler, M., Motta, G., and Von~Sachs, R. (2011).
\newblock Fitting dynamic factor models to non-stationary time series.
\newblock {\em Journal of Econometrics}, 163(1):51--70.

\bibitem[\protect\astroncite{Eickmeier et~al.}{2015}]{eickmeier2015classical}
Eickmeier, S., Lemke, W., and Marcellino, M. (2015).
\newblock Classical time varying factor-augmented vector autoregressive models
  estimation, forecasting and structural analysis.
\newblock {\em Journal of the Royal Statistical Society: Series A (Statistics
  in Society)}, 178(3):493--533.

\bibitem[\protect\astroncite{Ergemen et~al.}{2016}]{ergemen2016common}
Ergemen, Y.~E., Haldrup, N., and Rodr{\'\i}guez-Caballero, C.~V. (2016).
\newblock Common long-range dependence in a panel of hourly nord pool
  electricity prices and loads.
\newblock {\em Energy Economics}, 60:79--96.

\bibitem[\protect\astroncite{Forni et~al.}{2000}]{forni2000generalized}
Forni, M., Hallin, M., Lippi, M., and Reichlin, L. (2000).
\newblock The generalized dynamic-factor model: Identification and estimation.
\newblock {\em The Review of Economics and Statistics}, 82(4):540--554.

\bibitem[\protect\astroncite{Forni et~al.}{2004}]{forni2004generalized}
Forni, M., Hallin, M., Lippi, M., and Reichlin, L. (2004).
\newblock The generalized dynamic factor model consistency and rates.
\newblock {\em Journal of Econometrics}, 119(2):231--255.

\bibitem[\protect\astroncite{Forni et~al.}{2005}]{forni2005generalized}
Forni, M., Hallin, M., Lippi, M., and Reichlin, L. (2005).
\newblock The generalized dynamic factor model: one-sided estimation and
  forecasting.
\newblock {\em Journal of the American Statistical Association},
  100(471):830--840.

\bibitem[\protect\astroncite{Gao and Tsay}{2019}]{gao2019structural}
Gao, Z. and Tsay, R.~S. (2019).
\newblock A structural-factor approach to modeling high-dimensional time series
  and space-time data.
\newblock {\em Journal of Time Series Analysis}, 40(3):343--362.

\bibitem[\protect\astroncite{Garc{\i}a-Martos
  et~al.}{2012}]{garcia2012forecasting}
Garc{\i}a-Martos, C., Rodr{\i}guez, J., and Sanchez, M. (2012).
\newblock Forecasting electricity prices by extracting dynamic common factors:
  application to the iberian market.
\newblock {\em IET Generation, Transmission \& Distribution}, 6(1):11--20.

\bibitem[\protect\astroncite{Lam et~al.}{2012}]{lam2012factor}
Lam, C., Yao, Q., et~al. (2012).
\newblock Factor modeling for high-dimensional time series: inference for the
  number of factors.
\newblock {\em The Annals of Statistics}, 40(2):694--726.

\bibitem[\protect\astroncite{Maciejowska and
  Weron}{2015}]{maciejowska2015forecasting}
Maciejowska, K. and Weron, R. (2015).
\newblock Forecasting of daily electricity prices with factor models: utilizing
  intra-day and inter-zone relationships.
\newblock {\em Computational Statistics}, 30(3):805--819.

\bibitem[\protect\astroncite{Mikkelsen et~al.}{2018}]{mikkelsen2018consistent}
Mikkelsen, J.~G., Hillebrand, E., and Urga, G. (2018).
\newblock Consistent estimation of time-varying loadings in high-dimensional
  factor models.
\newblock {\em Journal of Econometrics}.

\bibitem[\protect\astroncite{Motta et~al.}{2011}]{motta2011locally}
Motta, G., Hafner, C.~M., and von Sachs, R. (2011).
\newblock Locally stationary factor models: Identification and nonparametric
  estimation.
\newblock {\em Econometric Theory}, 27(6):1279--1319.

\bibitem[\protect\astroncite{Pe{\~n}a and Box}{1987}]{pena1987identifying}
Pe{\~n}a, D. and Box, G.~E. (1987).
\newblock Identifying a simplifying structure in time series.
\newblock {\em Journal of the American statistical Association},
  82(399):836--843.

\bibitem[\protect\astroncite{Pe{\~n}a and
  Poncela}{2006}]{pena2006nonstationary}
Pe{\~n}a, D. and Poncela, P. (2006).
\newblock Nonstationary dynamic factor analysis.
\newblock {\em Journal of Statistical Planning and Inference},
  136(4):1237--1257.

\bibitem[\protect\astroncite{Poncela et~al.}{2021}]{poncela2021factor}
Poncela, P., Ruiz, E., and Miranda, K. (2021).
\newblock Factor extraction using kalman filter and smoothing: This is not just
  another survey.
\newblock {\em International Journal of Forecasting}.

\bibitem[\protect\astroncite{Porto et~al.}{2008}]{portomoreaub2008}
Porto, R., Morettin, P., and Aubin, E. C.~Q. (2008).
\newblock Wavelet regression with correlated errors on a piecewise holder
  class.
\newblock {\em Statistics \& Probability Letters}, 78(16):2739--2743.

\bibitem[\protect\astroncite{Rodr{\'\i}guez-Caballero and
  Ergemen}{2017}]{rodriguez2017estimation}
Rodr{\'\i}guez-Caballero, C.~V. and Ergemen, Y.~E. (2017).
\newblock Estimation of a dynamic multilevel factor model with possible
  long-range dependence.
\newblock Technical report, Universidad Carlos III de Madrid. Departamento de
  Estad{\'\i}stica.

\bibitem[\protect\astroncite{Rodr{\'\i}guez-Caballero
  et~al.}{2020}]{rodriguez2020air}
Rodr{\'\i}guez-Caballero, C.~V., Vera-Vald{\'e}s, J.~E., et~al. (2020).
\newblock Air pollution and mobility in the mexico city metropolitan area, what
  drives the covid-19 death toll?
\newblock Technical report, Department of Economics and Business Economics,
  Aarhus University.

\bibitem[\protect\astroncite{Stock and Watson}{2002}]{stock2002forecasting}
Stock, J.~H. and Watson, M.~W. (2002).
\newblock Forecasting using principal components from a large number of
  predictors.
\newblock {\em Journal of the American Statistical Association},
  97(460):1167--1179.

\bibitem[\protect\astroncite{Su and Wang}{2017}]{suwang2017}
Su, L. and Wang, X. (2017).
\newblock On time-varying factor models: Estimation and testing.
\newblock {\em Journal of Econometrics}, 198(1):84--101.

\bibitem[\protect\astroncite{Vidakovic}{2009}]{vidakovic2009statistical}
Vidakovic, B. (2009).
\newblock {\em Statistical modeling by wavelets}, volume 503.
\newblock John Wiley \& Sons.

\bibitem[\protect\astroncite{Weron}{2007}]{weron2007modeling}
Weron, R. (2007).
\newblock {\em Modeling and forecasting electricity loads and prices: A
  statistical approach}, volume 403.
\newblock John Wiley \& Sons.

\bibitem[\protect\astroncite{Weron}{2014}]{Weron20141030}
Weron, R. (2014).
\newblock Electricity price forecasting: A review of the state-of-the-art with
  a look into the future.
\newblock {\em International Journal of Forecasting}, 30(4):1030 -- 1081.

\end{thebibliography}

\clearpage

\appendix
\setcounter{table}{0} \setcounter{figure}{0}

\noindent {\Large {\textbf{Appendix}}}


\section{Technical appendix}
\subsection{Proof of Theorem \ref{propoCP}}\label{AP1}

\begin{proof}
	Let $\textbf{Y}=(Y_1,\ldots,Y_T)'$ be a $T\times N$ matrix, and let $V_{NT}$ be a $r\times r$ diagonal matrix composed by the $r$ largest eigenvalues of the matrix $(NT)^{-1}\textbf{Y}\textbf{Y}'$. By definition of eigenvectors and eigenvalues, we have
	\begin{equation}\label{CCPP} 
	\frac{1}{NT}\textbf{Y}\textbf{Y}'\tilde{\textbf{F}}=\tilde{\textbf{F}}V_{NT} \ \ \Longleftrightarrow  \ \  \frac{1}{NT}\textbf{Y}\textbf{Y}'\tilde{\textbf{F}}V_{NT}^{-1}=\tilde{\textbf{F}},
	\end{equation} 
	where $\tilde{\textbf{F}}'\tilde{\textbf{F}}=\mathbb{I}_r.$ 
	The model defined in (\ref{3Mod}) and (\ref{3Mod2}) can be re-written as
	\begin{equation*}
	\begin{array}{lll}
	\textbf{Y}_t & =  & [\boldsymbol{\Lambda}_0+\boldsymbol{\Lambda}(t)]\textbf{F}_t+\textbf{e}_t \\
	& =  & \boldsymbol{\Lambda}_0\textbf{F}_t+\boldsymbol{\Lambda}(t)\textbf{F}_t+\textbf{e}_t \\
	& =  & \boldsymbol{\Lambda}_0\textbf{F}_t+\textbf{w}_t+\textbf{e}_t,
	\end{array}
	\end{equation*}
	where $\textbf{w}_t=\boldsymbol{\Lambda}(t)\textbf{F}_t.$ Now, we define the following $T\times N$ matrices, $\textbf{e}=(\textbf{e}_1, \textbf{e}_2, \ldots, \textbf{e}_T)'$ and $\textbf{w}=(\textbf{w}_1, \textbf{w}_2, \ldots, \textbf{w}_T)'$. Note that the model in (\ref{3Mod}), can also be re-written in matrix form as
	\begin{equation*}\label{aux2}
	\textbf{Y}=\textbf{F}\boldsymbol{\Lambda}_0^{'}+\textbf{w}+\textbf{e}.
	\end{equation*}
	Consequently, after taking products we get
	\begin{equation*}\label{aux3}	
	\textbf{Y}\textbf{Y}'=\textbf{F}\boldsymbol{\Lambda}_0^{'}\boldsymbol{\Lambda}_0\textbf{F}'+\textbf{F}\boldsymbol{\Lambda}_0^{'}(\textbf{e}+\textbf{w})'+(\textbf{e}+\textbf{w})\boldsymbol{\Lambda}_0\textbf{F}'+(\textbf{e}+\textbf{w})(\textbf{e}+\textbf{w})'.
	\end{equation*}
	
	Then, from the definition of $\tilde{\textbf{F}}_t$ in (\ref{CCPP}) and the matrix rotation $H$, we can write for a fixed $t$
	\begin{equation*}\label{aux4}
	\begin{array}{lll}
	\vspace{0.25cm}
	\tilde{\textbf{F}}_t-H'\textbf{F}_t & =  & (NT)^{-1}V_{NT}^{-1}\tilde{\textbf{F}}'\textbf{Y}\textbf{Y}_t^{'}-V_{NT}^{-1}(\tilde{\textbf{F}}'\textbf{F}/T)(\boldsymbol{\Lambda}_0^{'}\boldsymbol{\Lambda}_0/N)\textbf{F}_t \\ 
	\vspace{0.25cm}

	&=&\frac{V_{NT}^{-1}}{NT}[\tilde{\textbf{F}}'(\textbf{F}\boldsymbol{\Lambda}_0^{'}\boldsymbol{\Lambda}_0\textbf{F}_t+\textbf{F}\boldsymbol{\Lambda}_0^{'}(w_t+e_t)+(w+e)\boldsymbol{\Lambda}_0\textbf{F}_t+(e+w)(e_t+w_t))\\
	\vspace{0.25cm}
	& &-(\tilde{\textbf{F}}'\textbf{F})(\boldsymbol{\Lambda}_0^{'}\boldsymbol{\Lambda}_0)\textbf{F}_t]\\
	\vspace{0.25cm}
	
	&=&\frac{V_{NT}^{-1}}{NT}[
	\underbrace{\tilde{\textbf{F}}'\textbf{F}\boldsymbol{\Lambda}_0^{'}e_t}_{D_{1t}}+\underbrace{\tilde{\textbf{F}}'e\boldsymbol{\Lambda}_0\textbf{F}_t}_{D_{2t}}+\underbrace{\tilde{\textbf{F}}'ee_t}_{D_{3t}}+\underbrace{\tilde{\textbf{F}}'\textbf{F}\boldsymbol{\Lambda}_0^{'}w_t}_{D_{4t}}+\underbrace{\tilde{\textbf{F}}'w\boldsymbol{\Lambda}_0\textbf{F}_t}_{D_{5t}}+\underbrace{\tilde{\textbf{F}}'ww_t}_{D_{6t}}+\\
	\vspace{0.25cm}
	&&
	\underbrace{\tilde{\textbf{F}}'ew_t}_{D_{7t}}+\underbrace{\tilde{\textbf{F}}'we_t}_{D_{8t}}]\\
	\vspace{0.25cm}
	&=&V_{NT}^{-1}\sum_{i=1}^{8}D_{it}.
	\end{array}
	\end{equation*}
		
	Then, after applying squared norms in both sides, adding on $t$, and dividing by $T$ in $V_{NT}^{-1}\sum_{i=1}^{8}D_{it}$, we get by the Löv inequality 
	\begin{equation}\label{aux6}
	\frac{1}{T}\sum_{t=1}^T \| \tilde{\textbf{F}}_t-H'\textbf{F}_t \|^2 \leq \|V_{NT}^{-1}\|^28 \sum_{i=1}^{8}  \left(\frac{1}{T}\sum_{t=1}^{T}\|D_{it}\|^2\right)
	\end{equation}
	where 
	\begin{equation*}
	\begin{array}{lll}
	D_{1t}=\tilde{\textbf{F}}'\textbf{F}\boldsymbol{\Lambda}_0^{'}e_t/{NT} &  &
	D_{2t}=\tilde{\textbf{F}}'e\boldsymbol{\Lambda}_0\textbf{F}_t/{NT}\\
	
	D_{3t}=\tilde{\textbf{F}}'ee_t/{NT} & &
	D_{4t}=\tilde{\textbf{F}}'\textbf{F}\boldsymbol{\Lambda}_0^{'}w_t/{NT}\\
	
	D_{5t}=\tilde{\textbf{F}}'w\boldsymbol{\Lambda}_0\textbf{F}_t/{NT} & &
	D_{6t}=\tilde{\textbf{F}}'ww_t/{NT}\\
	
	D_{7t}=\tilde{\textbf{F}}'ew_t/{NT}& &
	D_{8t}=\tilde{\textbf{F}}'we_t/{NT}\\
	\end{array}
	\end{equation*}
	
	Then, from \cite{mikkelsen2018consistent} (Lemma A.1), $V_{NT}$ converges to a definite positive matrix, therefore $\|V_{NT}^{-1}\|=O_{p}(1)$. Considering Assumptions A-E and properties of principal components, for each term $D_{it}, i=1,\ldots,8$, we have
	
	\begin{eqnarray*}
		T^{-1}\sum_{t=1}^{T}\|D_{1t}\|^2&=&O_p(N^{-1}),\\ T^{-1}\sum_{t=1}^{T}\|D_{2t}\|^2&=&O_p(N^{-1}T^{-1}),\\
		T^{-1}\sum_{t=1}^{T}\|D_{3t}\|^2&=&O_p(N^{-1}T^{-1}),\\  T^{-1}\sum_{t=1}^{T}\|D_{4t}\|^2&=&O_p(N^{-2}K_{1NT}), \\
		T^{-1}\sum_{t=1}^{T}\|D_{5t}\|^2&=&O_p(N^{-2}T^{-2}K_{2NT}),\\ T^{-1}\sum_{t=1}^{T}\|D_{6t}\|^2&=&O_p(N^{-2}T^{-2}K_{3NT}),\\
		T^{-1}\sum_{t=1}^{T}\|D_{7t}\|^2&=&O_p(N^{-2}K_{1NT}),\\ T^{-1}\sum_{t=1}^{T}\|D_{8t}\|^2&=&O_p(N^{-2}K_{1NT}).
	\end{eqnarray*}
	
	Finally, the right-hand side of equation (\ref{aux6}) is a sum of variables with orders
	$\left\{\frac{1}{N}, \frac{1}{NT}, \frac{K_{1NT}}{N^2}, \frac{K_{2NT}}{N^2T^2},\frac{K_{3NT}}{N^2T^2}\right\},$ respectively, and the proof is now completed.
\end{proof}

\clearpage

\section{Figures for Section \ref{sec:montecarlo}}

\begin{figure}[H]
	\centering
	\includegraphics[width=12.5cm,height=4cm]{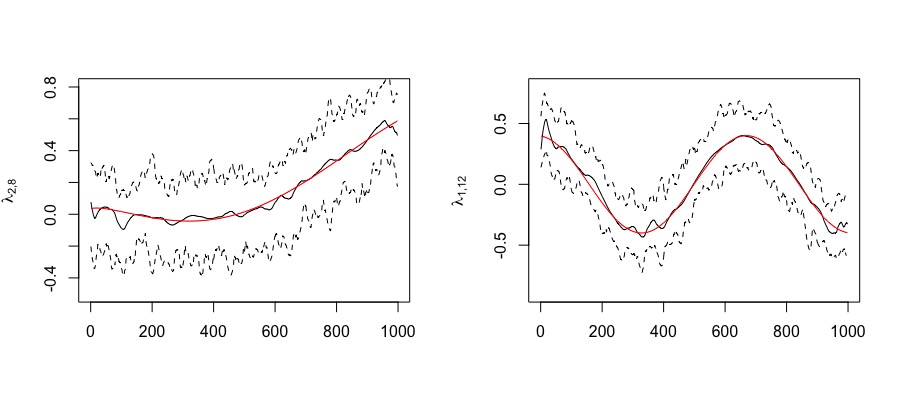}\\
	\includegraphics[width=12.5cm,height=4cm]{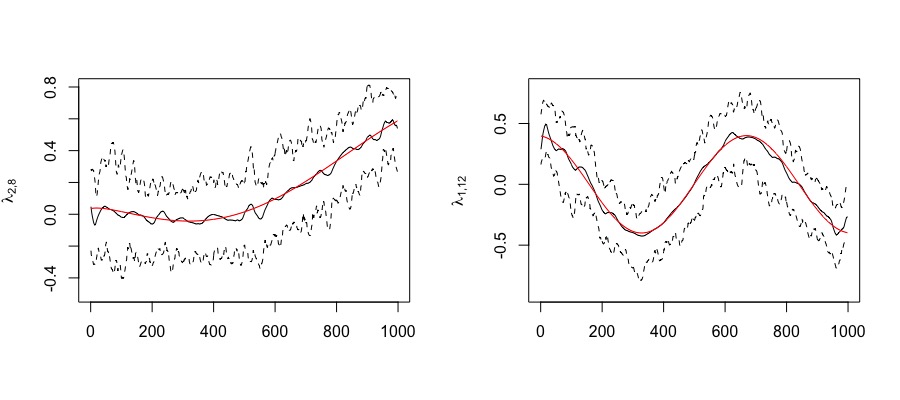}\\
	\includegraphics[width=12.5cm,height=4cm]{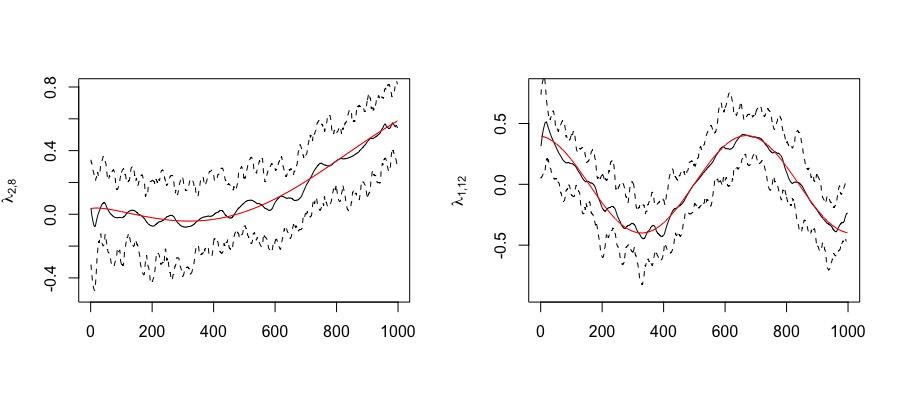}
	\caption{Comparison between the actual factor loadings (solid red line), the estimated factor loadings (solid black line), and Bootstrap confidence interval at 95\% (dashed black line). By column, from left to right: $\lambda_{1,12}$, and $\lambda_{28}$. By row, from top to
		bottom: $\theta_k=0$, $\theta_k=0.5$, and $\theta_k=1$, for $k \in \{1,2\}$. $\Gamma_e=Toep$, $N=20$, $T=1024$ and Wavelet Haar.}
	\label{loadingshaar}
\end{figure}

\begin{figure}[H]
	\centering
	\includegraphics[width=12.5cm,height=4cm]{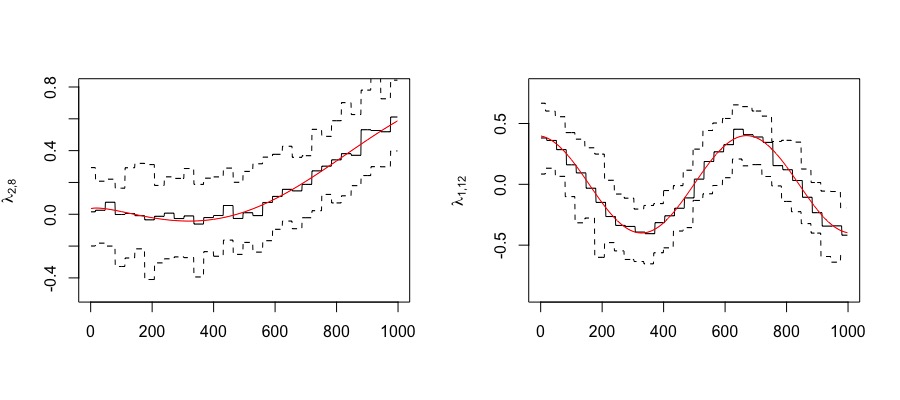}\\
	\includegraphics[width=12.5cm,height=4cm]{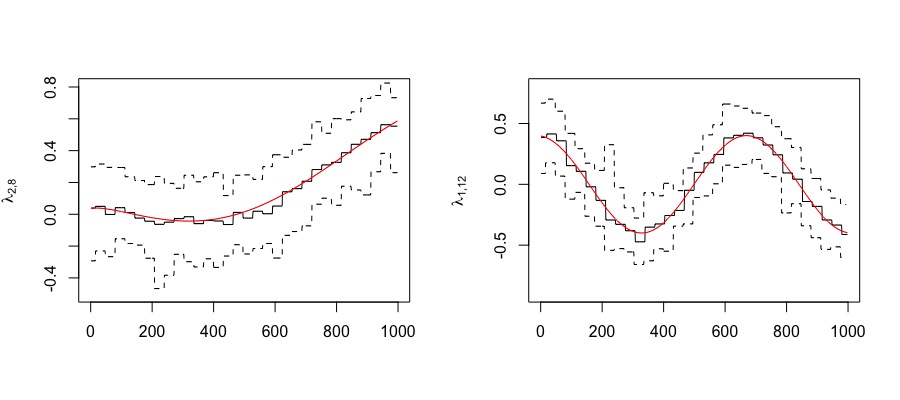}\\
	\includegraphics[width=12.5cm,height=4cm]{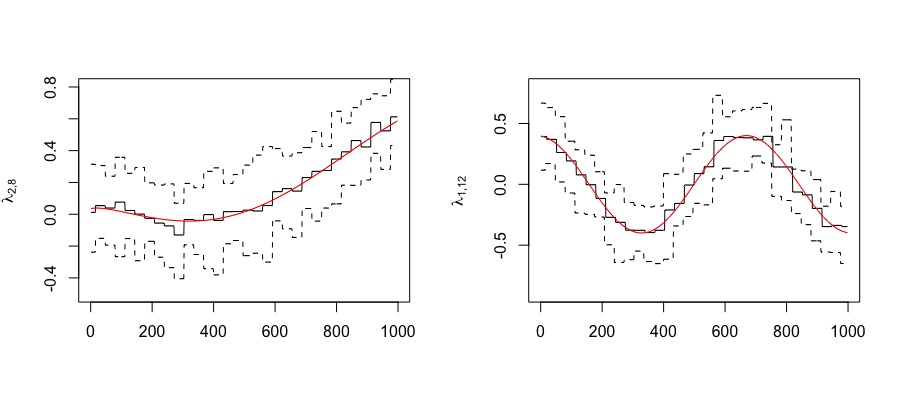}
	\caption{Comparison between the actual factor loadings (solid red line), the estimated factor loadings (solid black line), and Bootstrap confidence interval at 95\% (dashed black line). By column, from left to right: $\lambda_{1,12}$, and $\lambda_{28}$. By row, from top to
		bottom: $\theta_k=0$, $\theta_k=0.5$, and $\theta_k=1$, for $k \in \{1,2\}$. $\Gamma_e=Toep$, $N=20$, $T=1024$ and Wavelet D8.}
	\label{loadingsd8}
\end{figure}

\end{document}